\documentclass[pdflatex,sn-mathphys-num]{sn-jnl}
\usepackage[utf8]{inputenc}
\usepackage[T1]{fontenc}
\usepackage{graphicx}%
\usepackage{multirow}%
\usepackage{amsmath,amssymb,amsfonts}%
\usepackage{amsthm}%
\usepackage{mathrsfs}%
\usepackage[title]{appendix}%
\usepackage{xcolor}%
\usepackage{textcomp}%
\usepackage{manyfoot}%
\usepackage{booktabs}%
\usepackage{algorithm}%
\usepackage{algorithmicx}%
\usepackage{algpseudocode}%
\usepackage{listings}%

\theoremstyle{thmstyleone}%

\theoremstyle{thmstyletwo}%

\theoremstyle{thmstylethree}%

\begin{document}

\title{\centering
Dual Fear Mechanisms Shaping\\
Stochastic Population Dynamics\\
under the Allee Effect
}
\author*[1]{\fnm{Özgür} \sur{Gültekin}}\email{gultekino@yahoo.com}
\author[2]{\fnm{Mirza} \sur{Muradli}}

\affil*[1]{\orgname{Independent Researcher}, \orgaddress{\city{Kadıköy}, \state{İstanbul}, \country{Türkiye}}}
\affil[2]{\orgname{Üsküdar American Academy}, \city{Üsküdar}, \state{İstanbul}, \country{Türkiye}}

\abstract{Traditional population models that include predator--prey interactions attribute demographic changes directly to predation-related effects. However, predator-induced fear in prey has increasingly been recognised as an important factor shaping population dynamics. In this study, we propose a cubic population model in which fear acts through two distinct functional channels for a single-species population exhibiting the Allee effect. In this model, fear reduces the intrinsic growth rate through a multiplicative suppression mechanism while also playing an integrated role in modulating the growth and interaction dynamics by rescaling the saturation structure of the Holling type III interaction term. The stochastic extension of the model is described by a Langevin formalism containing correlated additive and multiplicative Gaussian noise, and the steady state probability distribution (SSPD) is analytically obtained using the corresponding Fokker--Planck equation. The analytical solution is validated by numerical simulations. The SSPD reveals both noise-induced transitions and fear-controlled regime changes between low- and high-density states, with the two-channel effect of fear producing structural competition and non-monotonic changes in the distribution. These are analysed through phenomenological bifurcation (P-bifurcation) diagrams and three-dimensional distribution surfaces. Additionally, statistical properties, parameter sensitivity, and escape dynamics are investigated through normalised moments, Fisher information, and mean first-passage time (MFPT) calculations. Notably, our model treats fear as an independent control parameter and provides a natural explanation for several conflicting empirical findings in the literature on fear-mediated population dynamics, while also offering an analytical basis for conservation biology and ecosystem management.}

\keywords{Fear effect, Allee effect, Stochastic population dynamics, Fokker--Planck equation, Steady state probability distribution, Noise-induced transitions, Regime shifts, Mean first-passage time.}

\maketitle
\clearpage
\section{Introduction}\label{sec:introduction}

Predator--prey population models based on interspecies interactions play a fundamental role in theoretical ecology and mathematical biology, as well as in complex systems, stochastic dynamics, and statistical physics \cite{murray2002,strogatz2019,roy2025communphys,mckane2005,dobramysl2018,fussmann2000,grunert2021,roy2025epjb,guin2023,moradi2015}. In classical Lotka--Volterra models, the main mechanism determining prey population size is the direct consumption of prey by the predator. However, in recent years, it has been shown that risk effects arising from predator presence are at least as decisive for prey population dynamics as direct consumption \cite{cresswell2011,hua2014,martin2011,lima1998nonlethal,preisser2008,lima1998stress}. The risk effect refers to the population-level demographic cost of the behavioural, morphological, and physiological responses that prey develop in response to predator presence \cite{sheriff2020}.

Independent studies conducted on different species and ecosystems have clearly demonstrated the decisive role of predator risk effects in population dynamics \cite{lima1998nonlethal,preisser2005,kats1998}. Zanette et al.~\cite{zanette2011} manipulated only the perception of fear by completely preventing direct predation in song sparrows (\textit{Melospiza melodia}) and showed that predator risk reduced annual offspring production by approximately 40\%. Schmitz et al.~\cite{schmitz1997} showed that, in experiments using risk-inducing spiders with glued jaws, grasshoppers that moved to lower-quality habitats solely in response to perceived predator presence could generate demographic losses equivalent to actual predation. Nelson et al.~\cite{nelson2004} showed that behavioural effects alone, without predator consumption, could suppress the population growth of pea aphids by approximately 30\%. Furthermore, it was found that this behavioural effect could constitute a significant portion of the total predator effect, in some cases approximately 80\%, depending on the experimental conditions.

Prey individuals that perceive predator presence spend more time vigilant, restrict their feeding activities, and often move to safer but food-poor habitats \cite{schmitz1997,brown2004}. The decrease in energy intake reduces individual condition, while elevated stress levels negatively affect reproductive success and the probability of survival \cite{clinchy2013,elliott2016}. The fear effect is not only short-term but can also have intergenerational demographic consequences \cite{allen2022}. It has been experimentally shown that offspring of parents exposed to predator fear exhibit lower performance due to developmental stress and have a reduced probability of survival \cite{elliott2016,benard2004}.

Moll et al.~\cite{moll2017} analysed the ways in which predator risk is defined within a hierarchical framework, revealing that a total of 244 different risk metrics were used across 141 studies. However, they found that much of the empirical literature focused on documenting the existence of the fear effect rather than investigating the functional interactions between these metrics, and that the questions of which mechanism fear operates through and at what scale it operates were largely left unanswered. In their evaluation based on the distinction between encounter rate $(a)$ and post-encounter mortality probability $(d)$ in Lima and Dill's~\cite{lima1990} formulation, they showed that the existing metrics mostly captured changes in encounter rate, whereas they represented the mortality-probability component to a much more limited extent \cite{moll2017}.

Some conflicting findings have been reported in the literature regarding the effect of predator risk on populations, which are consistent with the observations of Moll et al. For example, the effect of wolves on deer herd size in Yellowstone National Park has yielded inconsistent results across different studies. Some studies have shown that herd size increases as risk increases \cite{mao2005}, whereas others have shown that it decreases \cite{creel2005}, and still others have reported no significant effect \cite{gude2006}. These differences are thought to stem from the fact that risk is defined at different spatial and temporal scales and using different metrics \cite{moll2017}.

Such contradictory findings necessitate considering the structural vulnerabilities of population dynamics, especially in low-density regimes. The Allee effect \cite{courchamp2019}, defined as the positive relationship between population density and average individual fitness, makes low-density populations vulnerable to extinction. Difficulty finding mates at low densities, weakened group defences, and reduced social interactions eliminate the expected growth advantages \cite{courchamp2019}. Elliott et al.~\cite{elliott2017} experimentally demonstrated that predator fear can increase the risk of extinction by inducing an Allee effect in low-density populations. This finding suggests that the fear effect interacts with Allee dynamics.

With the increasing empirical evidence regarding the role of the fear effect in ecological systems, numerous theoretical population models incorporating the fear effect have been developed \cite{roy2020,sarkar2020,halder2020,kim2022,sk2023,biswas2024,bentout2021,chchaoui2025,yan2026}. Wang et al.~\cite{wang2016} were the first to incorporate the fear effect into population models as a multiplicative term that suppresses the intrinsic growth rate of the prey, independent of direct consumption. Expanding on this approach, Sasmal~\cite{sasmal2018} mathematically demonstrated that fear strengthens Allee dynamics in low-density regimes by considering fear and the Allee effect together. Subsequent studies have analysed the fear effect within models enriched with various mechanisms, such as time delays \cite{pal2023,parida2025}, prey refuges \cite{wang2019}, inter-predator cooperation \cite{umrao2025,roy2025jappl}, memory effects \cite{pal2023}, persistent fear costs \cite{li2026}, counterattack \cite{takyi2023,reshma2025,nguyen2025}, eco-epidemiological dynamics \cite{antwifordjour2024}, and non-local fear effects \cite{wang2025}.

In the literature, the fear parameter is usually included in models only as a suppressive factor acting on the intrinsic growth rate \cite{wang2016,sasmal2018}. In this study, we base our model on a single-species population model with the Allee effect and model predator-induced fear instead of centring the predator. We modulate the Holling type III term with the fear parameter and reinterpret it not only as a classical term representing the direct predation rate, but also as a nonlinear interaction term reflecting the combined effect of behavioural and encounter-based processes. Thus, we define a second interaction channel in our model and express the predation effect through an effective control parameter representing behavioural risk, without including the predator as an explicit dynamical variable. To our knowledge, this is the first time in the literature that a single fear parameter is proposed to be included in the model through two distinct mechanisms. Our approach proposes a general framework for multi-mechanism fear models, consistent with the multidimensional risk structure highlighted by Moll et al.~\cite{moll2017} and Lima and Dill's~\cite{lima1990} framework that distinguishes between encounter rate and post-encounter demographic outcomes. The fear parameter acting through two channels reveals a richer dynamic that determines the topology of the steady state distribution and regime organisation in the stochastic approach, while at the deterministic level it alters the position of the equilibrium points and the stability structure of the system. Furthermore, our model exhibits two different ecological scenarios depending on the sign of the interaction term. We introduce these scenarios in the next section.

Langevin and Fokker--Planck formalisms offer a suitable framework for the stochastic treatment of single-species population models \cite{gultekin2021,mandal2023,ghosh2012,yang2019}. We construct the stochastic representation of the model within a Langevin formalism that includes additive and multiplicative white noises and their correlation. We obtain the steady state solution analytically by deriving the Fokker--Planck equation and validating the results with numerical simulations performed using the Milstein method. Through parameter scans based on three-dimensional representations of the steady state distribution and P-bifurcation diagrams, we identify both noise-induced and fear-parameter-induced regime transitions and phase-regime shifts. Furthermore, our two-channel fear model generates non-monotonic regime dynamics in a stochastic environment and can explain Yellowstone-like contradictory empirical observations \cite{moll2017,mao2005,creel2005,gude2006} within the model's own dynamics. We quantitatively characterise the statistical structure of the distribution using normalised moments and Fisher information analysis. Finally, we derive a Kramers-type mean first-passage time expression to describe the escape dynamics from the high-density steady state and reveal the non-monotonic behaviour of transition times as functions of noise intensity, correlation, and the fear parameter.

This study, beyond its ecological consequences, offers more general physical implications for regime organisation in stochastic systems with multi-channel control parameters. Unlike the fixed barrier structure predicted by the classical Kramers escape framework, the two-channel control mechanism dynamically reshapes the effective potential landscape and extends the transition dynamics beyond purely diffusion-driven processes.

The study is structured as follows. In Section~\ref{sec:model}, the deterministic model involving the two-channel fear mechanism and the two ecological scenarios is introduced. In Section~\ref{sec:sspd}, the steady state probability distribution is obtained via the Fokker--Planck formalism and validated with numerical simulations. In Section~\ref{sec:pbifurcation}, P-bifurcation analysis is presented, and in Section~\ref{sec:fear-regime}, the role of the fear parameter in stochastic regime organisation is examined. In Section~\ref{sec:statistical}, the statistical properties of the distribution are characterised through normalised moments and Fisher information. In Section~\ref{sec:mfpt}, Kramers-type mean first-passage time analysis is performed. In Section~\ref{sec:conclusions}, the results are presented.

\section{Model}\label{sec:model}

For a single-species prey population whose dynamics are determined by the Allee effect, we propose a cubic population model that includes the fear effect both as a stress factor that multiplicatively suppresses the intrinsic growth rate and as a saturation modulator that upper-bounds the demographic output dependent on predator--prey contact through the Holling type III term:
\begin{equation}
\frac{dx}{dt}
=
rx\left(1-\frac{x}{K}\right)(x-m)\frac{1}{1+fv}
-
\frac{a_1x^2}{1+a_2x^2}.
\label{eq:model}
\end{equation}

Here, \(x(t)\) represents the average population size at time \(t\); \(r\), the intrinsic growth rate of the population; \(K\), the environmental carrying capacity; \(m\), the Allee threshold; \(f\), the level of fear developed by prey in response to the predator; and \(v\), the predator population density. The model includes a weak Allee effect when \(m<0\) and a strong Allee effect, due to the additional root that appears on the positive side, when \(m>0\). The population is driven towards extinction when its size falls below the Allee threshold value \(m\). Due to the biological meanings of the parameters, all parameters except \(m\) and \(a_1\) are defined as positive. When the modelling preference \(a_2=f\) is adopted, the fear parameter \(f\) functions not only as a multiplier that suppresses the intrinsic growth rate, but also as an integrated control parameter that simultaneously rescales the saturation structure of the Holling type III term. The model describes two different scenarios depending on the sign of \(a_1\).

\subsection*{Scenario I: \(a_1>0\)}

The fear parameter, which scales intrinsic growth, represents the demographic cost of the prey's stress response to predator presence and is reflected in the model as a suppression channel that monotonically reduces the prey population. In the contact channel, the Holling type III term represents the demographic loss due to predator--prey contact. With the modelling choice \(a_2=f\), fear gains a second function as a modulation parameter that upper-bounds the magnitude of this contact-mediated loss through the saturation structure. Ecologically, in this scenario, as the fear level increases, prey develop risk-avoidance behaviours, and consequently, the increase in predation pressure is limited by the saturation structure of the contact term. Since the Holling type III term is rescaled by fear, it can now be interpreted as a contact-mediated prey consumption term representing the demographic output of behavioural consumption pressure, rather than directly representing predation loss.

Thus, fear produces structural competition by operating simultaneously through two opposing dynamical channels based on the mechanisms of (i) suppression of growth and (ii) saturation-limited contact-mediated consumption. This competition determines the position of fixed points and the stability structure in the deterministic model. In the stochastic approach, however, it gives rise to a much richer dynamical structure.

\subsection*{Scenario II: \(a_1<0\)}

In this scenario, the Holling type III term represents a contact channel that partially damps the fear-induced pressure in the growth channel through the prey's risk-avoidance behaviours under predator pressure. When \(a_1<0\), this term does not add new individuals to the population; instead, it reduces the net effect of negative pressure in the growth channel through the contact channel. With the modelling choice \(a_2=f\), fear also upper-bounds this damping capacity. As fear increases, the effectiveness of the contact channel decreases, and its capacity to balance pressure in the growth channel progressively weakens.

Thus, in this scenario, fear also produces structural competition by operating simultaneously through two opposing dynamical channels based on the mechanisms of (i) suppression of growth and (ii) limitation of the damping capacity of the contact channel.

In both scenarios, in the absence of a predator \((v=0)\), the parameter representing the intensity of the predator--prey contact effect is taken as \(a_1=0\), and thus the model reduces to the Allee-effect growth term for a single species. In the presence of a predator, in the limit \(f\to0\), the fear multiplier in the growth channel becomes \(1/(1+fv)\to1\), and the Holling type III term reduces to a contact term of the form \(-a_1x^2/(1+fx^2)\to -a_1x^2\). This boundary condition represents the baseline condition in Scenario I, where there is no fear-induced growth pressure and the contact channel produces demographic loss at maximum efficiency. In Scenario II, it represents the baseline condition where there is no fear-induced growth pressure and no fear-induced saturation mechanism.

In the presence of a predator, in the high-fear limit \(f\to\infty\), the fear multiplier in the growth channel is damped as \(1/(1+fv)\to0\), and the Holling type III term is damped as \(-a_1x^2/(1+fx^2)\to0\). In Scenario I, this regime represents a significant demographic decrease caused by stress on the intrinsic growth rate and a situation where contact-induced loss is completely suppressed by the saturation mechanism. In Scenario II, both the fear suppression in the growth channel reaches its maximum and the damping capacity of the contact channel completely disappears. Consequently, in the high-fear regime, the net demographic output of the contact channel approaches zero at the boundary for both scenarios, although the underlying mechanisms and biological interpretations differ.

\begin{figure}[h]
\centering
\includegraphics[width=1\textwidth]{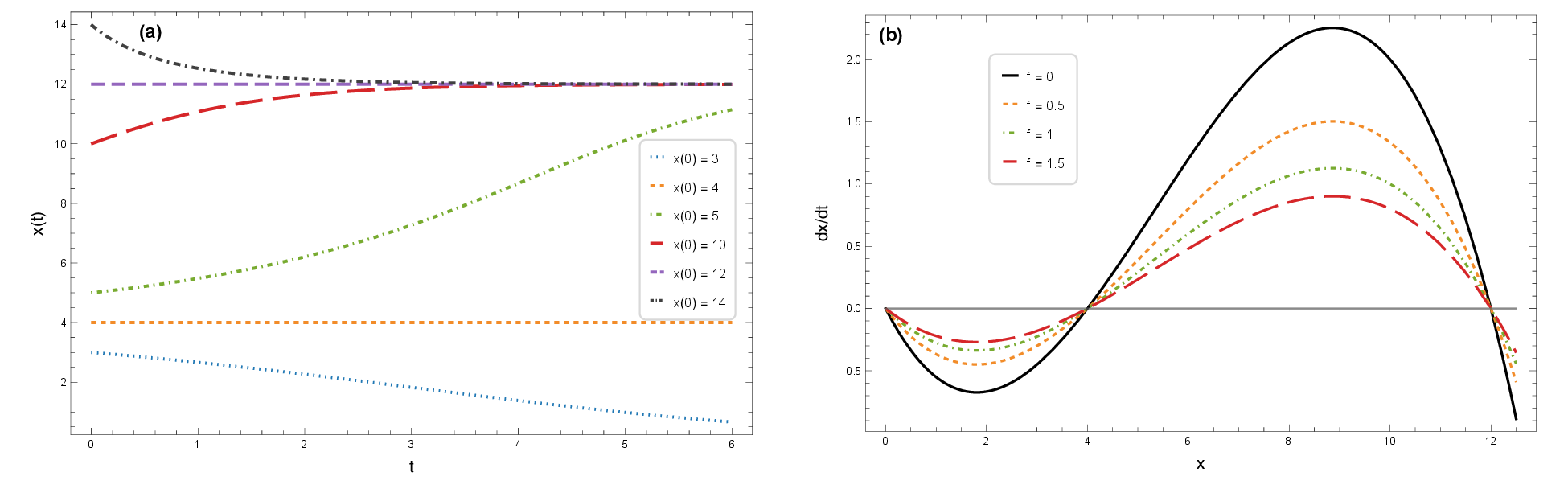}
\vspace{-0.8em}
\caption{For the parameters \(r=0.2\), \(K=12\), \(m=4\), \(v=1\), and \(a_1=0\): (a) average population size \(x(t)\) as a function of time for different initial conditions, with \(f=0.5\); (b) time derivative of the average population size, \(dx/dt\), as a function of \(x\) for different \(f\) values.}
\label{fig:fig1}
\end{figure}

The fundamental novelty of our model is that fear is treated as an integrated control parameter acting simultaneously through two separate functional channels. If \(a_1=0\) is taken, fear is reduced to a stress factor that only suppresses the growth rate, as in previous studies. In this case, the rich nonlinear dynamical structure arising from the dual-channel interaction is not observed. As seen in Figure~\ref{fig:fig1}, even at a non-zero fear level \((f=0.5)\), in the strong Allee-effect regime, if the initial population is above the Allee threshold \((x(0)>m)\), all trajectories converge to the carrying capacity \(K\), while if the initial population is below the threshold \((x(0)<m)\), the population is driven towards extinction. Figure~\ref{fig:fig1} also shows that the fear level does not change the equilibrium points of the system. The stable equilibrium points of the system remain \(x=0\) and \(x=K\), and these are independent of fear.

\vspace{-0.5em}

\begin{figure}[h]
\centering
\includegraphics[width=1\textwidth]{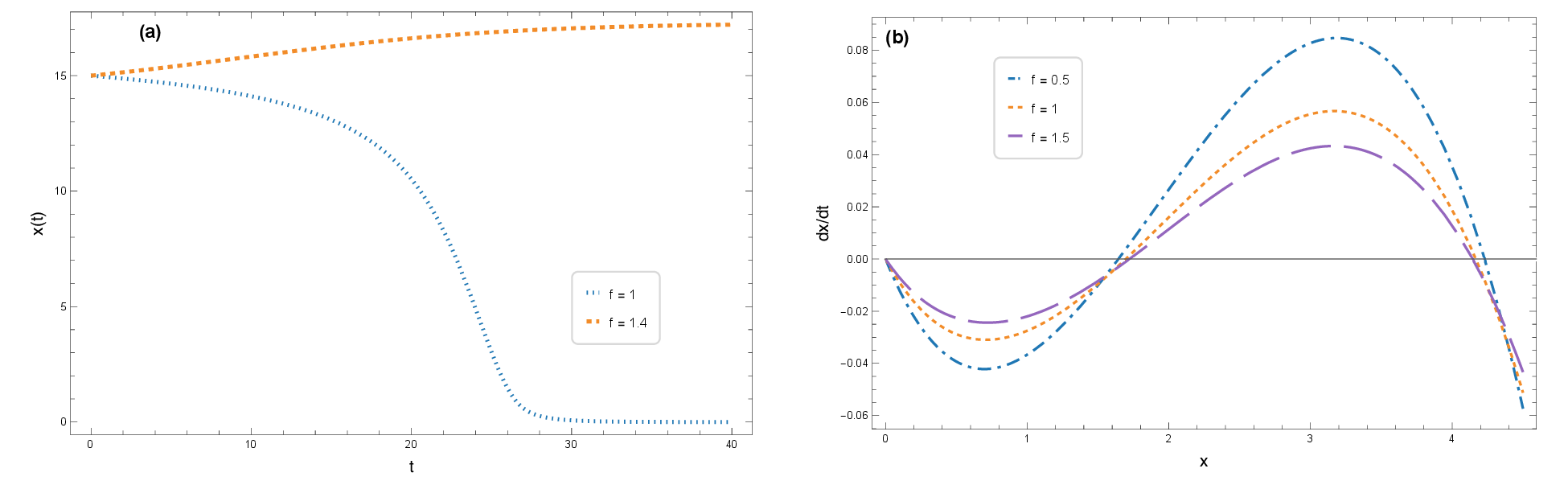}
\vspace{-1em}
\caption{For the parameters \(r=0.1\), \(v=1\), and \(a_2=f\): (a) average population size \(x(t)\) as a function of time for \(f=1\) and \(f=1.4\) under the same initial condition, with \(K=20\), \(m=10\), \(x(0)=15\), and \(a_1=1\); (b) time derivative of the average population size, \(dx/dt\), as a function of \(x\) for different \(f\) values, with \(K=4\), \(m=2\), and \(a_1=-0.02\).}
\label{fig:fig2}
\end{figure}

Figure~\ref{fig:fig2} clearly shows the distinctive feature of our model. Trajectories with the same initial condition are directed to different equilibrium points for different values of the fear parameter. In Figure~\ref{fig:fig2}, for \(f=1\), the population converges to its carrying capacity, whereas for \(f=1.4\), it moves towards extinction. This effect arises because fear not only suppresses growth but also rescales the demographic output of predator contact. Furthermore, Figure~\ref{fig:fig2} shows that the equilibrium points of the system change with fear. In other words, the bidirectional role of the fear parameter in the growth and contact channels provides a mechanism that reorganises the velocity-field topology and equilibrium structure of the system depending on the intensity of fear.

\vspace{-0.5em}

To stochastically describe our model, we write the following Langevin equation:
\begin{equation}
\frac{dx}{dt}
=
rx\left(1-\frac{x}{K}\right)(x-m)\frac{1}{1+fv}
-
\frac{a_1x^2}{1+a_2x^2}
+
x\Gamma(t)
+
\eta(t).
\label{eq:langevin}
\end{equation}

Here, \(\eta(t)\) represents zero-mean Gaussian additive white noise, and \(\Gamma(t)\) represents multiplicative white noise. The correlation properties of the noises are defined as

\begin{equation}
\begin{aligned}
\langle \eta(t)\eta(t')\rangle &= 2\alpha \delta(t-t'),\\
\langle \Gamma(t)\Gamma(t')\rangle &= 2\beta \delta(t-t'),\\
\langle \eta(t)\Gamma(t')\rangle &= 2\gamma\sqrt{\alpha\beta}\,\delta(t-t').
\end{aligned}
\label{eq:noise_correlations}
\end{equation}

Here, \(\alpha\) and \(\beta\) express the additive and multiplicative noise intensities, respectively, and \(\gamma\) expresses the degree of correlation between the two noises.

The deterministic potential corresponding to the model is given by
\begin{equation}
V(x)
=
\frac{1}{1+fv}
\left[
\frac{r}{4K}x^4
-
\frac{r(m+K)}{3K}x^3
+
\frac{mr}{2}x^2
+
\frac{a_1(1+fv)}{a_2}x
\right]
-
\frac{a_1}{a_2^{3/2}}
\arctan\left(\sqrt{a_2}x\right).
\label{eq:deterministic_potential}
\end{equation}
\vspace{-1em}
\begin{figure}[h]
\centering
\includegraphics[width=0.65\textwidth]{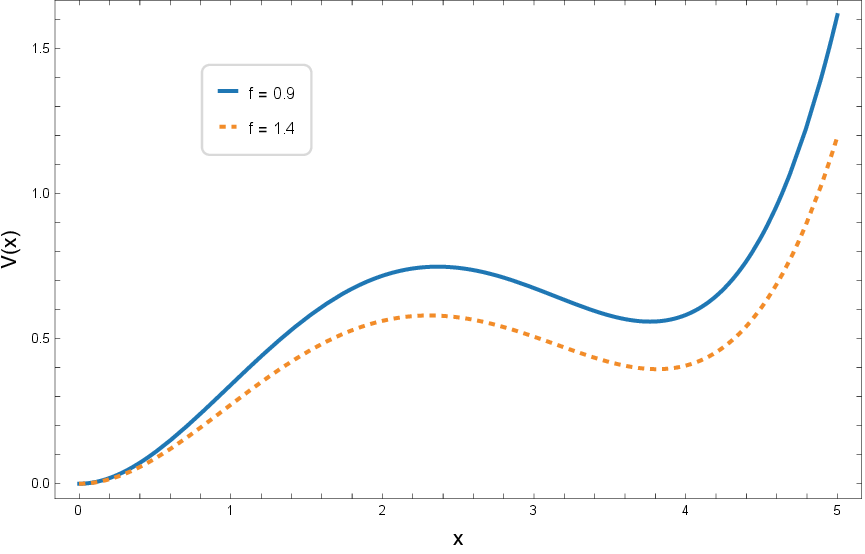}
\caption{Deterministic potential \(V(x)\) as a function of \(x\) for the parameters \(r=3\), \(K=4\), \(m=2\), \(v=1\), \(a_1=0.2\), and \(a_2=f\).}
\label{fig:fig3}
\end{figure}
\vspace{-0.5em}

The deterministic potential in Eq.~\eqref{eq:deterministic_potential} is plotted in Figure~\ref{fig:fig3} by taking \(a_2=f\). Figure~\ref{fig:fig3} clearly shows how fear intensity affects the equilibrium structure of the population. In the potential landscape, the well depth, barrier height, and positions of the extrema change depending on fear intensity.

\section{Steady state probability distribution}\label{sec:sspd}

In this section, we derive the Fokker--Planck equation associated with the Langevin equation in Eq.~\eqref{eq:langevin} and obtain the steady state probability distribution of the system:
\begin{equation}
\frac{\partial P(x,t)}{\partial t}
=
-\frac{\partial}{\partial x}\left[A(x)P(x,t)\right]
+
\frac{\partial^2}{\partial x^2}\left[B(x)P(x,t)\right].
\label{eq:fokker_planck}
\end{equation}

Here, \(P(x,t)\) is the probability density function of the population size at time \(t\). Using the statistical properties of the noise terms defined in Eq.~\eqref{eq:noise_correlations}, the drift term \(A(x)\) and the diffusion coefficient \(B(x)\) are obtained as \cite{dajin1994,gardiner1997,risken1996}
\begin{equation}
A(x)
=
\frac{1}{1+fv}
\left[
-\frac{r}{K}x^3
+
r\left(1+\frac{m}{K}\right)x^2
-
mrx
\right]
-
\frac{a_1x^2}{1+a_2x^2}
+
\beta x
+
\gamma\sqrt{\alpha\beta},
\label{eq:drift}
\end{equation}
and
\begin{equation}
B(x)
=
\beta x^2
+
2\gamma\sqrt{\alpha\beta}\,x
+
\alpha.
\label{eq:diffusion}
\end{equation}
\vspace{0.5em}

In the steady state, by setting \(\partial P(x,t)/\partial t=0\) and imposing natural boundary conditions, the steady state probability density of the population size is obtained as follows \cite{gardiner1997,jacobs2010}:
\vspace{0.5em}
\begin{equation}
P_{\mathrm{st}}(x)
=
\frac{N}{B(x)}
\exp\left[
\int_{0}^{x}
\frac{A(x')}{B(x')}\,dx'
\right].
\label{eq:sspd_general}
\end{equation}
\vspace{0.5em}

Here, \(N\) is the normalisation constant, determined by the condition
\[
\int_{}^{x} P_{\mathrm{st}}(x)\,dx=1.
\]

When Eqs.~\eqref{eq:drift} and \eqref{eq:diffusion} are substituted into Eq.~\eqref{eq:sspd_general}, the integral
\begin{equation}
\Lambda(x)
=
\int^{x}
\frac{A(x')}{B(x')}\,dx'
\label{eq:lambda_definition}
\end{equation}
can be evaluated analytically. Thus, the steady state probability distribution can be written as
\begin{equation}
P_{\mathrm{st}}(x)
=
\frac{N e^{\Lambda(x)}}
{\beta x^2+2\gamma\sqrt{\alpha\beta}\,x+\alpha}.
\label{eq:sspd_compact}
\end{equation}
\vspace{0.5em}

However, since the explicit expression for \(\Lambda(x)\) is complicated, the full expression of the steady state probability density \(P_{\mathrm{st}}(x)\) is presented in the Appendix.

\begin{figure}[h]
\centering
\includegraphics[width=0.65\textwidth]{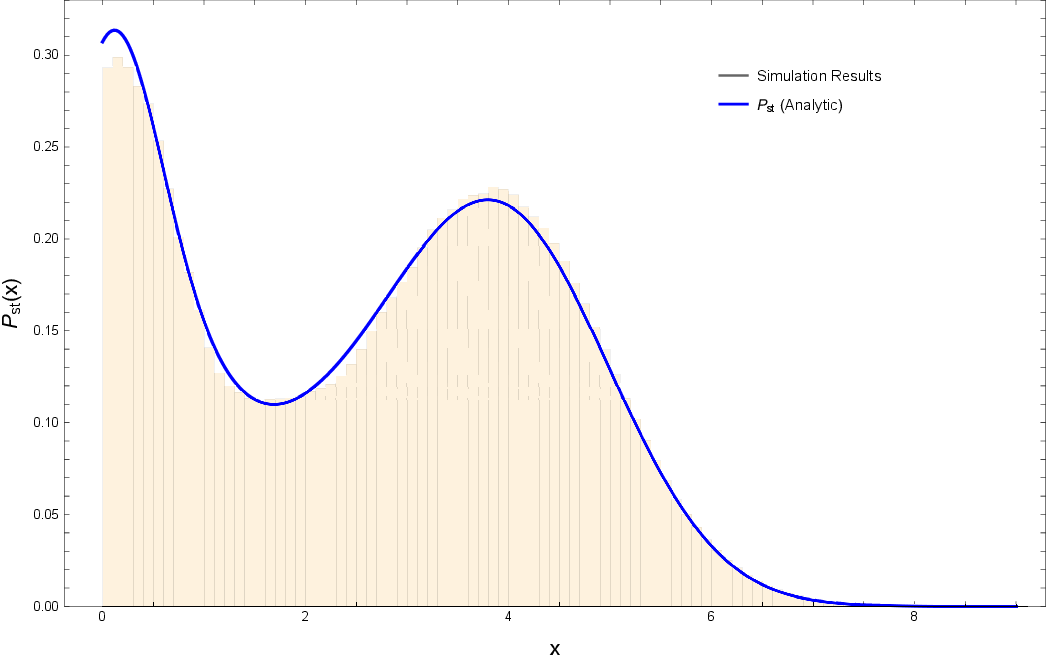}
\caption{Comparison between the analytically obtained steady state distribution \(P_{\mathrm{st}}(x)\) and the stationary distribution histogram obtained from Milstein simulations for the parameters \(r=1\), \(K=3\), \(m=2\), \(v=1\), \(f=0\), \(a_1=-0.2\), \(a_2=10^{-6}\), \(\alpha=0.73\), \(\beta=0.35\), and \(\gamma=-0.5\), with \(\Delta t=5\times10^{-5}\) and \(T=8000\).}
\label{fig:fig4}
\end{figure}

A numerical simulation was performed to test the accuracy of the analytically obtained steady state probability density \(P_{\mathrm{st}}(x)\) in Eq.~\eqref{eq:sspd_compact}. Taking into account the presence of multiplicative noise, the Milstein method \cite{kloeden1992} was used for numerical integration, and integration was performed for a total time of \(T=8000\) units with a time step of \(\Delta t=5\times10^{-5}\). After eliminating the initial transient regime, the stationary distribution was calculated from the obtained samples using the histogram method. As seen in Figure~\ref{fig:fig4}, very good agreement was observed between the numerical simulation results and the analytical solution.

\begin{figure}[h]
\centering
\includegraphics[
  width=1\textwidth,
  trim=17pt 0pt 9pt 0pt,
  clip
]{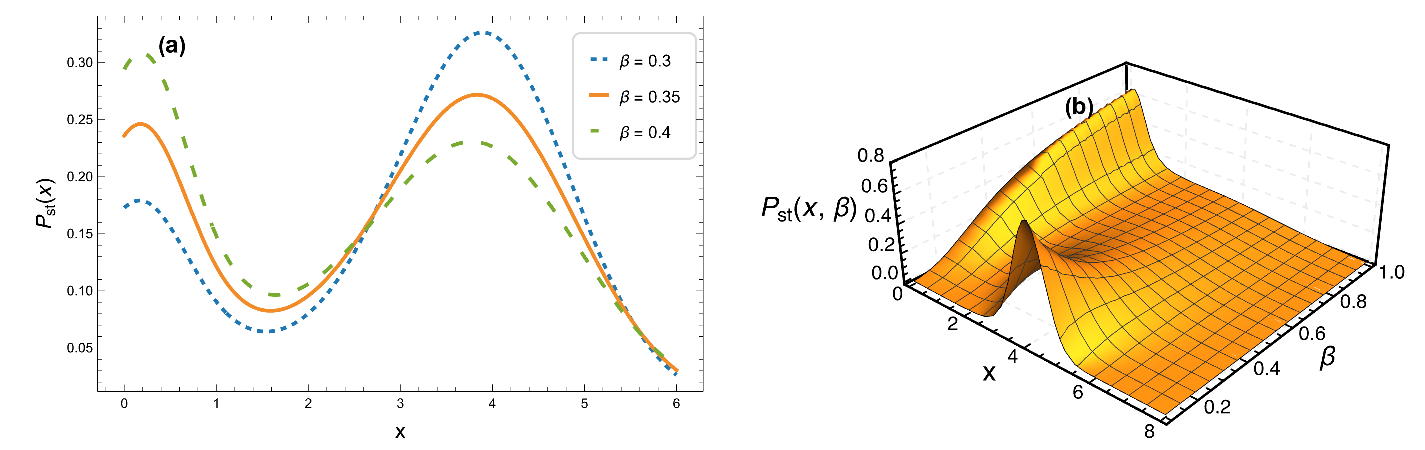}
\vspace{-0.5em}
\caption{For the parameters \(r=1\), \(K=3\), \(m=2\), \(v=1\), \(f=0\), \(\gamma=-0.7\), \(\alpha=0.75\), \(a_1=-0.2\), and \(a_2=10^{-6}\): (a) steady state distribution \(P_{\mathrm{st}}(x)\) as a function of population density \(x\) for different \(\beta\) values; (b) three-dimensional representation of \(P_{\mathrm{st}}(x,\beta)\) as a function of population density \(x\) and parameter \(\beta\).}
\label{fig:fig5}
\end{figure}

\begin{figure}[h]
\centering
\includegraphics[
  width=1\textwidth,
  trim=17pt 0pt 8pt 0pt,
  clip
]{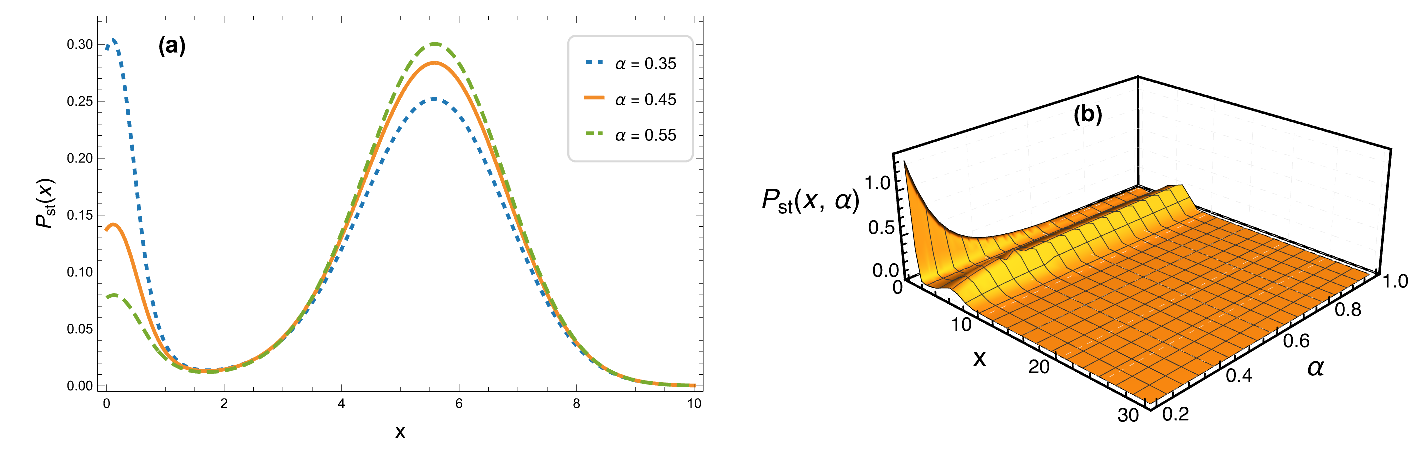}
\vspace{-0.5em}
\caption{For the parameters \(r=1\), \(K=4.5\), \(m=2\), \(v=1\), \(f=0\), \(\gamma=-0.65\), \(\beta=0.3\), \(a_1=-0.2\), and \(a_2=10^{-6}\): (a) steady state distribution \(P_{\mathrm{st}}(x)\) as a function of population density \(x\) for different \(\alpha\) values; (b) three-dimensional representation of \(P_{\mathrm{st}}(x,\alpha)\) as a function of population density \(x\) and parameter \(\alpha\).}
\label{fig:fig6}
\end{figure}

Figures~\ref{fig:fig5} and \ref{fig:fig6} show how the steady state probability density function \(P_{\mathrm{st}}(x)\) of the system changes depending on the noise parameters. The distribution exhibits a double-peaked structure in the examined parameter range. The relative heights of these peaks determine the effective behaviour of the system over long time scales. In Figure~\ref{fig:fig5}, the fact that the peak around the carrying capacity is higher for \(\beta=0.3\) shows that the effective behaviour of the system is determined by the high-density (HD) regime. In contrast, at \(\beta=0.4\), the effective behaviour of the system shifts to the low-density (LD) regime as the low-density peak becomes higher than the other peak. In Figure~\ref{fig:fig6}, it is seen that the system shifts from the low-density (LD) regime to the high-density (HD) regime as the additive noise intensity \(\alpha\) increases over the examined parameter values. Changes in noise parameters cause noise-induced regime transitions due to the rearrangement of probability weights in the stationary distribution.

Figures~\ref{fig:fig5} and \ref{fig:fig6} also include three-dimensional representations showing the evolution of the steady state probability distribution along \(\beta\) and \(\alpha\), respectively. The main difficulty in constructing these parameter-dependent three-dimensional surfaces is that, since the normalisation coefficient of the distribution explicitly depends on the noise parameters, \(P_{\mathrm{st}}(x)\) cannot be directly defined as a surface using a single closed expression. Therefore, \(P_{\mathrm{st}}(x)\) is normalised separately for each noise value, and the distribution is represented as a continuous surface along the parameter using the resulting parameter-dependent normalisation coefficients. This approach makes it possible to clearly observe how noise-induced regime transitions occur in parameter space between low-density (LD) and high-density (HD) regimes.

\section{P-bifurcation analysis}\label{sec:pbifurcation}

The number of extrema of a steady state probability distribution is directly related to the number of peaks of the distribution. For example, a double-peaked steady state probability distribution has a total of three extrema, consisting of two maxima and one minimum. Depending on the system parameters, the disappearance of one of these peaks leads to a decrease in the number of extrema. Therefore, examining how the extrema change with the system parameters is important for determining noise-induced regime changes. Such diagrams, which show the change in the extrema of the steady state probability distribution with parameters in a stochastic system, are called phenomenological bifurcation (P-bifurcation) diagrams \cite{arnold1995}.

To determine the extrema of the stationary distribution, we impose the condition \(dP_{\mathrm{st}}(x)/dx=0\). Using Eq.~\eqref{eq:sspd_general}, this condition can be written as \(B'(x)=A(x)\). Therefore, the extrema of \(P_{\mathrm{st}}(x)\) are determined by
\begin{equation}
\frac{r}{1+fv}
\left[
\frac{x^3}{K}
-
\left(1+\frac{m}{K}\right)x^2
+
mx
\right]
+
\frac{a_1x^2}{1+a_2x^2}
+
\beta x
+
\gamma\sqrt{\alpha\beta}
=0.
\label{eq:pbifurcation_roots}
\end{equation}

The non-negative roots of Eq.~\eqref{eq:pbifurcation_roots} represent the stationary states of the population.
\vspace{-1em}
\begin{figure}[h]
\centering
\includegraphics[width=1\textwidth]{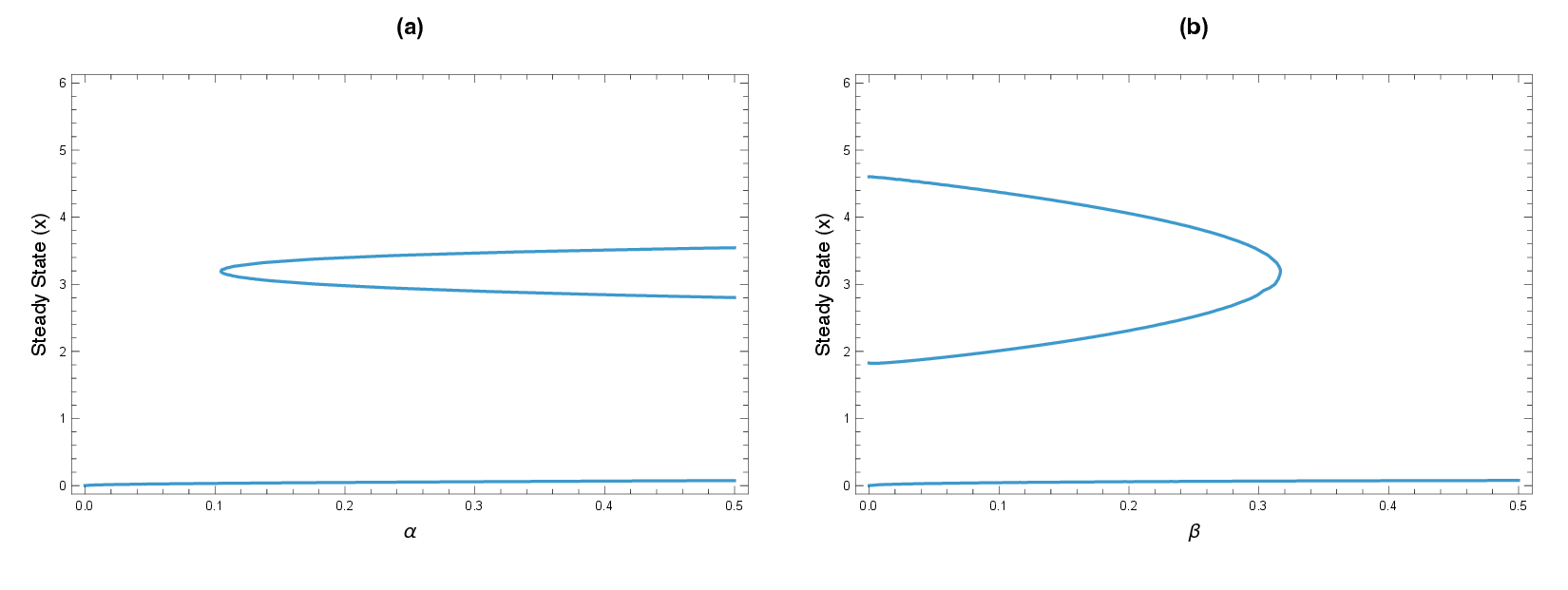}
\vspace{-1.5em}
\caption{P-bifurcation diagrams for the parameters \(r=1\), \(K=4.5\), \(m=2\), \(v=1\), \(\gamma=-0.3\), \(a_1=-0.1\), \(a_2=f\), and \(f=0.5\). (a) Along the additive noise intensity \(\alpha\) with \(\beta=0.3\) fixed; (b) along the multiplicative noise intensity \(\beta\) with \(\alpha=0.4\) fixed.}
\label{fig:fig7}
\end{figure}

The P-bifurcation diagrams presented in Figure~\ref{fig:fig7} show the variation in the number of extrema in the steady state probability distribution depending on the additive noise intensity \(\alpha\) and the multiplicative noise intensity \(\beta\). The emergence of more than one stationary solution at certain noise intensities along the \(\alpha\) axis in Figure~\ref{fig:fig7}a and along the \(\beta\) axis in Figure~\ref{fig:fig7}b corresponds to a multiple-extrema structure. P-bifurcation diagrams allow the number and locations of extrema to be directly tracked for both noise types. Since the number of extrema is related to the number of peaks, the branch mergers or disappearances observed at the parameter values where bifurcation occurs indicate qualitative changes in the peak structure.

\section{The role of fear in stochastic regime organisation}\label{sec:fear-regime}

Local minima of the deterministic potential correspond to stable equilibrium points, while local maxima correspond to unstable equilibrium points. However, in a stochastic environment, the long-time-scale behaviour of the system cannot be explained solely by this deterministic structure. The noise terms and the fear parameter \(f\), which acts on the system through two distinct functional channels, reorganise the depth of the potential wells and the positions of the extrema. Thus, the fear parameter directly controls the topological structure of the steady state probability distribution \(P_{\mathrm{st}}(x)\) and the probability transfer between these peaks.

Two different changes observed in \(P_{\mathrm{st}}(x)\) are distinguished. When the number of peaks of the stationary distribution remains constant while one peak becomes dominant over another, this is defined as a regime shift. In this case, the long-time-scale behaviour of the system shifts between low-density (LD) and high-density (HD) regimes. The number of extrema of the distribution is preserved, while only the probability weights are redistributed. On the other hand, a change in the number of peaks of \(P_{\mathrm{st}}(x)\), that is, a qualitative transformation of the extrema structure, is called a phase-regime shift. This change in the number of peaks can be clearly observed through P-bifurcation diagrams.

In the literature, qualitative changes in stochastic systems are mostly discussed within the framework of noise-induced transitions depending on noise intensity \cite{lefever2006,berglund2006}. However, in the model we propose, the transition mechanism is not limited to the noise parameters \((\alpha,\beta,\gamma)\). The fear parameter \(f\) emerges as an independent control parameter that determines the phase structure of the system by reorganising potential barriers and well depths, and can lead to parameter-induced transitions independent of noise.

\vspace{-0.9em}
\begin{figure}[H]
\centering
\includegraphics[width=0.88\textwidth]{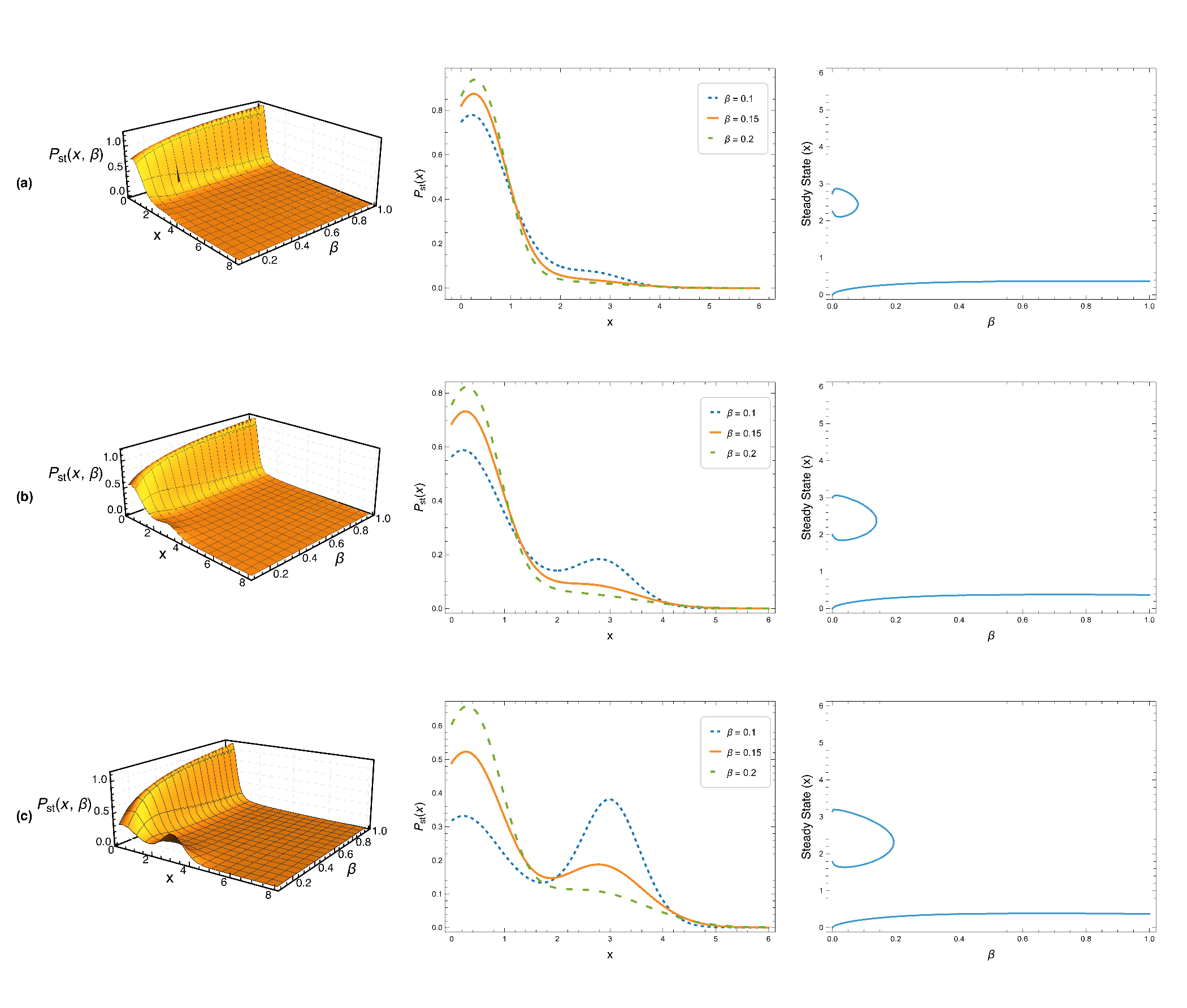}
\caption{The steady state probability distribution \(P_{\mathrm{st}}(x)\) as a function of \(x\) (middle panels), the three-dimensional representation of \(P_{\mathrm{st}}(x,\beta)\) (left panels), and the corresponding P-bifurcation diagrams (right panels) for the parameters \(r=1\), \(K=3\), \(m=2\), \(v=1\), \(f=1\), \(\alpha=0.5\), \(\gamma=-0.9\), and \(a_2=f\), with (a) \(a_1=0.1\), (b) \(a_1=0\), and (c) \(a_1=-0.1\).}
\label{fig:fig8}
\end{figure}

Figure~\ref{fig:fig8} presents the steady state probability distributions \(P_{\mathrm{st}}(x)\) for different \(\beta\) values, their three-dimensional representations, and the corresponding P-bifurcation diagrams. Figure~\ref{fig:fig8}c shows that for \(\beta=0.1\), the distribution exhibits a distinctly double-peaked structure. At \(\beta=0.15\), the distribution retains its double-peaked character, but the peak in the high-density (HD) region is relatively weaker. At \(\beta=0.1\), the peak in the high-density (HD) region is more dominant than the peak in the low-density (LD) region, while at \(\beta=0.15\) this is reversed. This corresponds to a noise-induced regime shift. At \(\beta=0.2\), the peak in the high-density region disappears, and the distribution becomes single-peaked. Here, a phase-regime shift occurs. This situation is clearly observed in the corresponding P-bifurcation diagram. For \(\beta=0.1\), there are three roots around \(x\approx0\), \(x\approx2\), and \(x\approx3\). The roots around \(x\approx0\) and \(x\approx3\) correspond to the local maxima of \(P_{\mathrm{st}}(x)\), while the root around \(x\approx2\) corresponds to its local minimum. The system has two stable and one unstable equilibrium point. In the P-bifurcation diagram, a single root is observed at \(\beta=0.2\), and this is consistent with the single-peaked structure of the distribution. Furthermore, the three-dimensional representation of \(P_{\mathrm{st}}(x)\) shows how this transition occurs along \(\beta\) and how the distribution transforms from a double-peaked to a single-peaked structure.

In Figure~\ref{fig:fig8}a, since \(a_1=0.1\), this corresponds to Scenario I, and the Holling type III term represents demographic loss due to contact. In this scenario, multiplicative noise \(\beta\) increases the fluctuations in the growth term, making the high-density (HD) regime, already suppressed by contact, more fragile. Therefore, the HD well disappears at a smaller \(\beta\) value compared to the cases in Figures~\ref{fig:fig8}b and \ref{fig:fig8}c. In other words, the system transitions to the low-density (LD) regime at lower \(\beta\) levels. In Figure~\ref{fig:fig8}b, since \(a_1=0\), the contact channel disappears, and the increase in \(\beta\) directly affects the growth term. In this case, the disappearance of the HD peak occurs at a larger \(\beta\) value compared to Scenario I, and the double-peaked structure is preserved over a wider \(\beta\) range than in Scenario I. In Figure~\ref{fig:fig8}c, since \(a_1=-0.1\), this corresponds to Scenario II. In this scenario, the Holling type III term represents a limited demographic contribution originating from the contact channel, and this contribution structurally supports the HD regime. Consequently, as multiplicative noise increases, the double-peaked structure is preserved over the widest \(\beta\) range, and the critical \(\beta\) value at which the HD well disappears is larger than in the other scenarios.

\vspace{-1.4em}
\begin{figure}[H]
\centering
\includegraphics[width=0.88\textwidth]{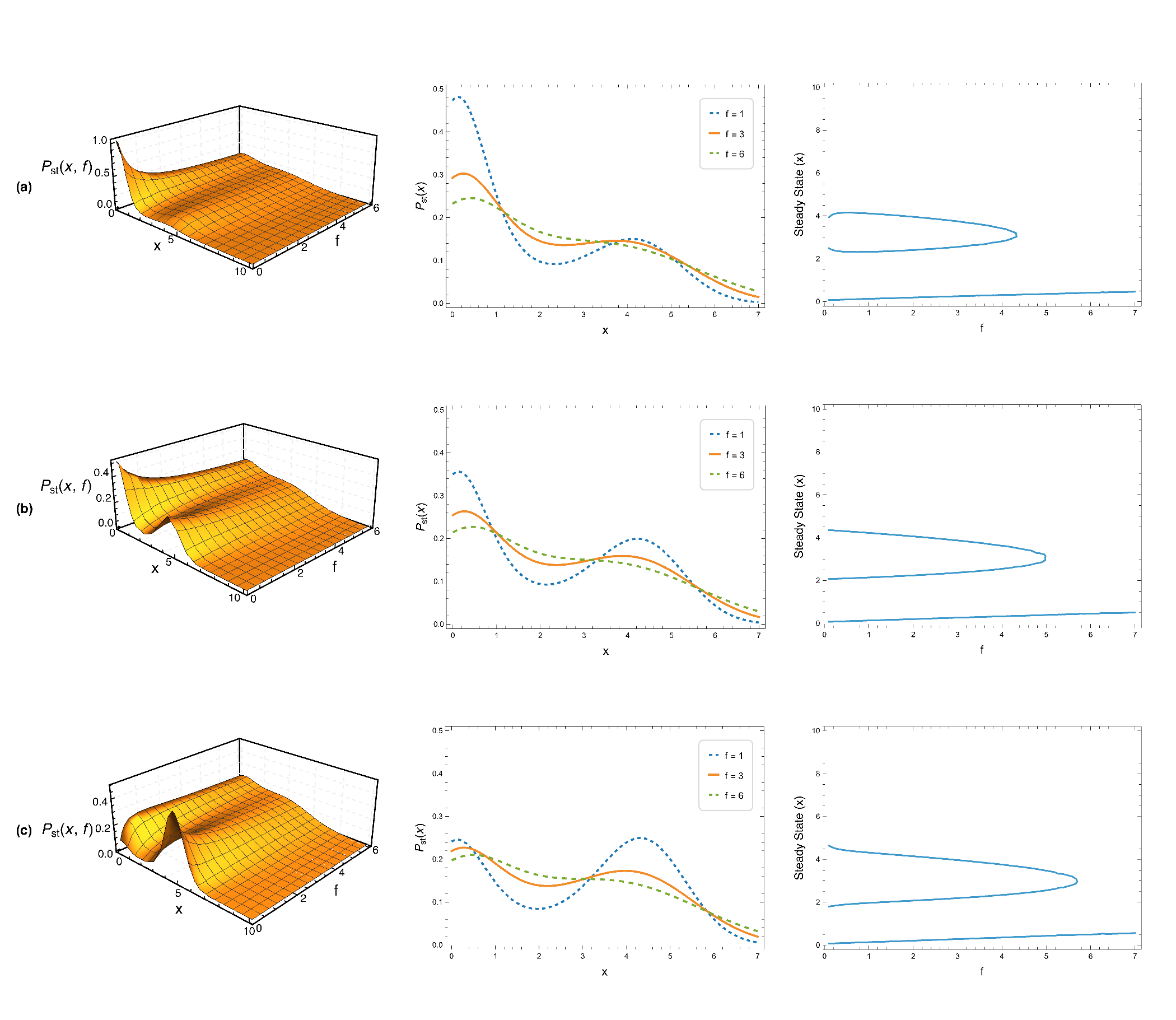}
\vspace{-0.5em}
\caption{The steady state probability distribution \(P_{\mathrm{st}}(x)\) as a function of \(x\) (middle panels), the three-dimensional representation of \(P_{\mathrm{st}}(x,f)\) (left panels), and the corresponding P-bifurcation diagrams (right panels) for the parameters \(r=1\), \(K=4.5\), \(m=2\), \(v=1\), \(\alpha=0.5\), \(\beta=0.1\), \(\gamma=-0.6\), and \(a_2=f\), with (a) \(a_1=0.1\), (b) \(a_1=0\), and (c) \(a_1=-0.1\).}
\label{fig:fig9}
\end{figure}

Figure~\ref{fig:fig9} presents the steady state probability distributions \(P_{\mathrm{st}}(x)\) for different \(f\) values, their three-dimensional representations, and the corresponding P-bifurcation diagrams. In this analysis, the noise parameters were kept constant, and only \(f\) was changed. Therefore, the changes observed in the structure of the distribution are directly parameter-driven. As seen in Figures~\ref{fig:fig9}a--\ref{fig:fig9}c, as \(f\) increases, the peaks in both the low-density (LD) and high-density (HD) regions generally decrease in height. In contrast, the probability density in the intermediate-density band increases, and the distribution evolves into a flatter form. This behaviour differs from the regime shifts observed in the \(\beta\)-scan, which are characterised by the transfer of peak weights from one regime to another. The fundamental behaviour observed in the \(f\)-scan is the simultaneous weakening of the double peak and the redistribution of probability into the intermediate-density region. In all figures, when \(f\) becomes sufficiently large, at a critical \(f\) value, the HD peak disappears, and the distribution becomes single-peaked. This transformation is followed in the P-bifurcation diagrams in a manner consistent with the decrease in the number of roots. While this trend is preserved in all three scenarios, the \(f\)-interval over which the double-peaked distribution can be maintained varies quantitatively depending on the scenario.

\begin{figure}[h]
\centering
\includegraphics[
  width=1\textwidth,
  trim=16pt 0pt 8pt 0pt,
  clip
]{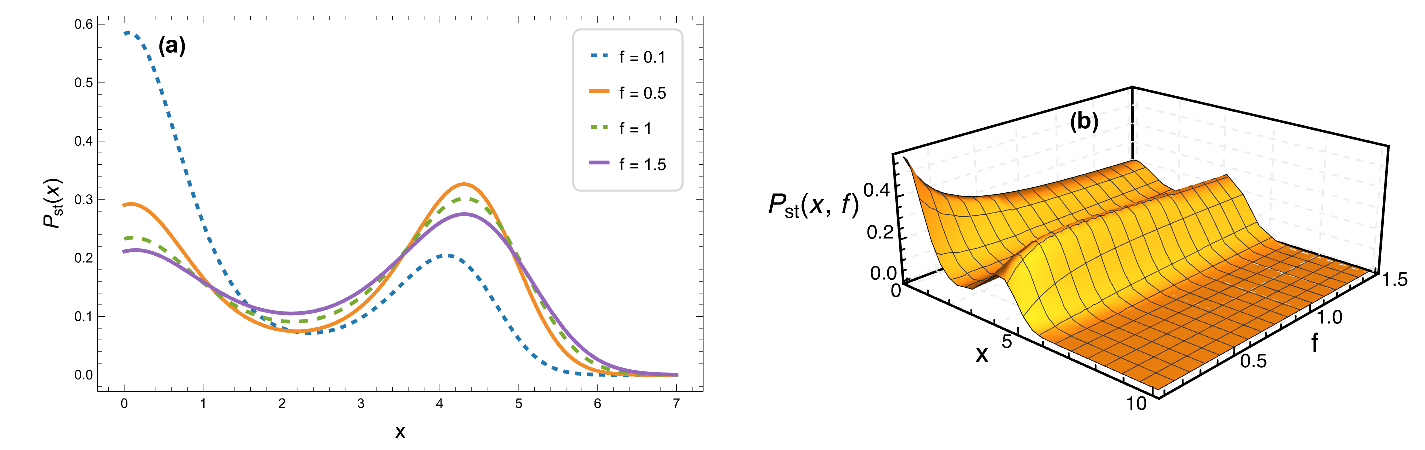}
\vspace{-0.8em}
\caption{The steady state probability distribution \(P_{\mathrm{st}}(x)\) as a function of \(x\) (a), and the three-dimensional representation of \(P_{\mathrm{st}}(x,f)\) as a function of population density \(x\) and the fear parameter \(f\) (b), for the parameters \(r=1\), \(K=4.5\), \(m=2\), \(v=1\), \(\alpha=0.7\), \(\beta=0.05\), \(\gamma=-0.6\), \(a_1=0.1\), and \(a_2=f\).}
\label{fig:fig10}
\end{figure}
\vspace{-0.3em}

The effect of the fear parameter \(f\) on the steady state probability distribution \(P_{\mathrm{st}}(x)\) under a fixed parameter set \((a_1=0.1,\text{ Scenario I})\) is shown in Figure~\ref{fig:fig10}. At a low fear level \((f=0.1)\), the distribution is double-peaked, with the peak in the low-density (LD) region being higher. At a higher fear value \((f=0.5)\), the relative height of the high-density (HD) peak increases, while the height of the low-density (LD) peak decreases. However, as \(f\) increases further \((f=1)\), the height of both peaks decreases overall, while the probability density in the intermediate-density band increases, and the distribution evolves into a flatter and more spread-out form. These observations show that the peak heights in the \(f\)-scan do not exhibit a monotonic change and that the process cannot be explained by a simple one-way peak transfer. At low \(f\) values, the high-density (HD) peak increases with increasing \(f\); however, at high \(f\) values, both the high-density (HD) and low-density (LD) peaks decrease simultaneously. This behaviour points to a more complex change compared to typical noise-induced transitions, which are frequently observed across noise parameters and characterised by the shifting of peak weights from one regime to another.

Studies on elk herds in Yellowstone National Park show that the effect of predator risk on population density is not one-way. It has been reported that increasing risk increases population density in some cases \cite{mao2005} and decreases it in others \cite{creel2005}. These contradictory results are thought to stem from the definition of predator risk at different temporal and spatial scales \cite{moll2017}. Our proposed dual-channel fear model has the potential to directly explain both the increase and decrease of peaks with increasing \(f\) in Figure~\ref{fig:fig10}, within the framework of the stochastic dynamics of the model. Therefore, \(f\) acts not only as a parameter that weakens the double-peaked character of the distribution, but also as a control parameter that reorganises the relative distribution of the probability mass between the LD and HD regions at certain intervals.

\section{Statistical properties of the steady state probability distribution}\label{sec:statistical}

In this section, we perform two complementary analyses to examine the quantitative effects of changes in noise intensities and the fear parameter on the steady state probability distribution (SSPD). First, we present the parameter-dependent variation of the normalised moments characterising the distribution. Second, we calculate Fisher information as an indicator measuring the parameter sensitivity of the distribution and examine its behaviour across the relevant parameters. These analyses quantitatively reveal the statistical structure of the SSPD and its parameter-dependent rearrangement dynamics.

\subsection{Moment analysis}\label{subsec:moment-analysis}

The \(n\)th-order moment of the population density \(x\) is given by
\begin{equation}
\langle x^n\rangle
=
\int_{0}^{\infty}x^nP_{\mathrm{st}}(x)\,dx.
\label{eq:nth_moment}
\end{equation}
The mean population density is defined as \(M_x \equiv E_x=\langle x\rangle\). Since the mean of the distribution changes significantly with respect to the parameters, we use normalised statistical concepts to reflect the scale-independent properties of the distribution \cite{xu2020}. The normalised variance is defined as
\begin{equation}
\mathrm{Var}_x
=
\frac{\langle x^2\rangle}{E_x^2}
-
1.
\label{eq:normalised_variance}
\end{equation}

In fact, this expression is equivalent to the square of the coefficient of variation, written as \(\mathrm{V}_x=\mathrm{Var}(x)/(E_x)^2\). Thus, simple scale effects resulting from parameter changes are eliminated. To evaluate the asymmetric nature of the distribution within the same scale-consistent framework, the normalised third moment is defined as
\begin{equation}
S_x
=
\frac{\langle x^3\rangle}{E_x^3}
-
3\mathrm{V}_x
-
1.
\label{eq:normalised_third_moment}
\end{equation}

Equation~\eqref{eq:normalised_third_moment} is equivalent to the expression \(S_x=\langle (x-E_x)^3\rangle/(E_x)^3\). Here, normalisation is performed with respect to the mean instead of the standard deviation, unlike the classical definition of skewness. This approach allows the comparison of the second and third moments with respect to the characteristic mean population scale of the system.

\begin{figure}[H]
\centering
\includegraphics[width=0.87\textwidth]{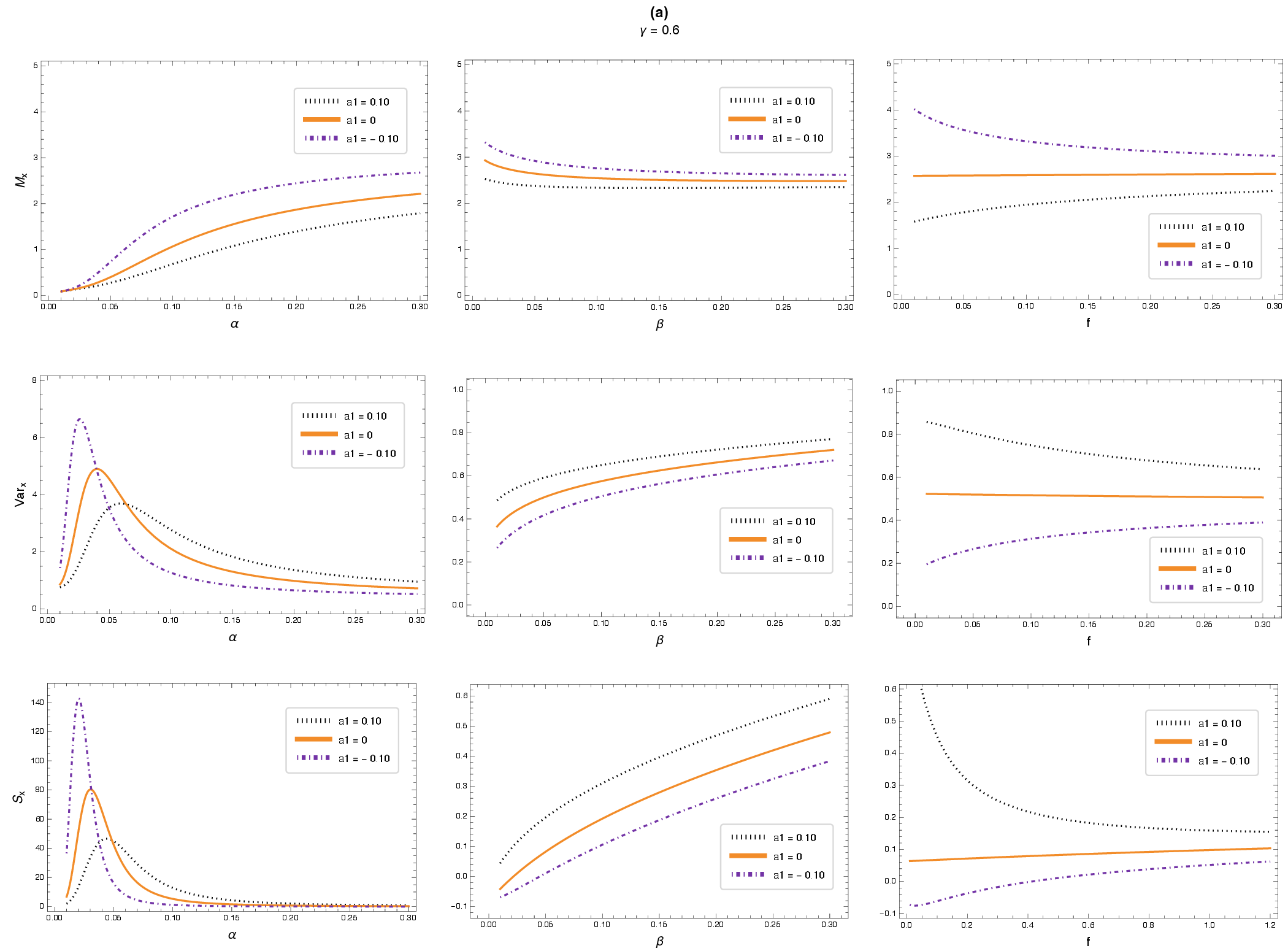}

\vspace{0.2em}

\hspace{0.6em}\includegraphics[width=0.87\textwidth]{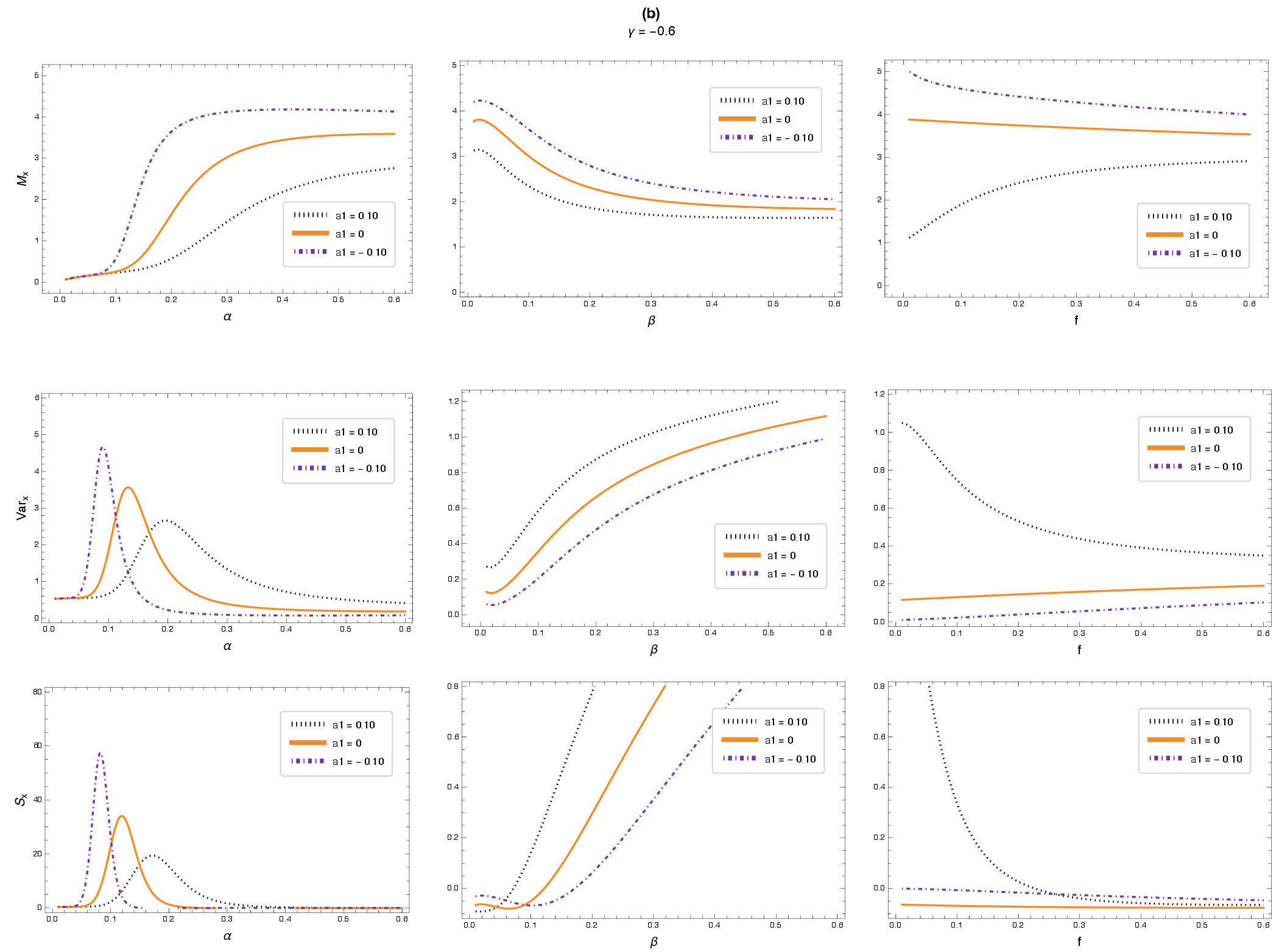}
\caption{Variation of the normalised moments of the steady state probability distribution under systematic parameter scans. The rows represent the normalised mean \(M_x\), normalised variance \(\mathrm{Var}_x\), and normalised third moment \(S_x\), respectively. Panel (a) corresponds to the correlation case \(\gamma=0.6\), and panel (b) corresponds to the correlation case \(\gamma=-0.6\). In both panels, \(r=1\), \(m=2\), \(K=4.5\), \(v=1\), and \(a_2=f\). In the first column, systematic scanning was performed along the \(\alpha\) parameter by fixing \(\beta=0.05\) and \(f=0.5\); in the second column, along the \(\beta\) parameter by fixing \(\alpha=0.7\) and \(f=0.5\); and in the third column, along the \(f\) parameter by fixing \(\alpha=0.7\) and \(\beta=0.05\).}
\label{fig:fig11}
\end{figure}

Figure~\ref{fig:fig11} shows the variation of the normalised moments \((M_x,\mathrm{Var}_x,S_x)\) along the additive noise intensity \(\alpha\), multiplicative noise intensity \(\beta\), and fear parameter \(f\). Figure~\ref{fig:fig11}a corresponds to the positive correlation case \((\gamma=0.6)\), and Figure~\ref{fig:fig11}b corresponds to the negative correlation case \((\gamma=-0.6)\).

For both positive and negative correlation, along \(\alpha\) in the first column, \(M_x\) increases monotonically from small \(\alpha\) values and approaches saturation at larger \(\alpha\) values. In contrast, \(\mathrm{Var}_x\) and \(S_x\) exhibit a distinct maximum in a narrow range at small \(\alpha\) values and then rapidly decay.

This distinct peak structure observed along the additive noise intensity \(\alpha\) is not observed along the multiplicative noise intensity \(\beta\). This situation can be explained by the asymmetric contributions arising from the diffusion coefficient structure
\(B(x)=\beta x^2+2\gamma\sqrt{\alpha\beta}\,x+\alpha.\) Since the multiplicative noise intensity \(\beta\) contributes to diffusion proportionally to \(x^2\), it remains relatively weak in the low-density region. In contrast, in the same region, the additive noise intensity \(\alpha\) has a more decisive effect as it contributes to the diffusion term independently of density. At small \(\alpha\) values, the total diffusion decreases to relatively small values, especially in the low-density region, and the distribution structure and the probability weights between regimes may show high sensitivity to \(\alpha\). This situation can contribute to the appearance of distinct peaks in variance and skewness. As \(\alpha\) increases, this sensitivity decreases, and the peaks quickly fade. Furthermore, it is observed that in the case of positive correlation, these peaks systematically reach higher values compared to the case of negative correlation. This finding indicates that the inter-noise correlation directly affects the parametric sensitivity of the distribution. However, this peak structure does not only stem from the shape of the diffusion coefficient. While the drift term \(A(x)\) determines the steady state structure and regime organisation of the system, the diffusion coefficient \(B(x)\) shapes which regions become statistically dominant on this structure and where the parametric sensitivity is concentrated. In addition, the fact that the peak values are higher in the \(a_1=-0.1\) scenario compared to the other scenarios can be attributed to the fact that, in this scenario, the damping capacity of the contact channel partially balances the growth pressure, making the probability mass more sensitive to parametric changes.

Along \(\beta\) in the second column, \(M_x\) tends to decrease overall in both correlation cases and for all scenarios. In contrast, \(\mathrm{Var}_x\) increases monotonically with increasing \(\beta\). Although \(S_x\) also tends to increase overall, in the case of negative correlation, at small \(\beta\) values, it first decreases to negative values and then rapidly increases again. This behaviour suggests that the opposite-phase noise coupling temporarily reverses the asymmetry of the distribution and that the increase in \(\beta\) not only increases the dispersion but also qualitatively changes the geometry of the distribution by rearranging the probability weights between regimes.

The behaviour of the moments along the fear parameter \(f\) in the third column is clearly dependent on the sign of \(a_1\). In the case of positive correlation \((\gamma=0.6)\), \(M_x\) remains approximately constant for \(a_1=0\), while it shows an increase for \(a_1=0.1\) and a decrease for \(a_1=-0.1\). On the other hand, in \(\mathrm{Var}_x\) and \(S_x\), a decrease is observed in the \(a_1=0.1\) scenario, and an increase is observed in the \(a_1=-0.1\) scenario. Under negative correlation \((\gamma=-0.6)\), these distinctions are generally preserved, but the quantitative levels change. When \(S_x\) is examined specifically, it is seen that in the case of positive correlation, especially in the \(a_1=-0.1\) scenario, \(S_x\) starts from negative values and increases, changing its sign to positive values with increasing \(f\). In contrast, under negative correlation, \(S_x\) remains negative in the \(a_1=0\) scenario, and in the other scenarios, it tends to move towards the negative region with increasing \(f\).

The fact that the behaviour of moments depending on \(f\) is so sensitive to the sign of \(a_1\) shows that the fear parameter operates as a guiding control mechanism in the system. In the case of \(a_1=0.1\), the increasing fear effect tends to reduce the relative amplitude of population fluctuations, contributing to the distribution being directed towards a narrower structure, while in the case of \(a_1=-0.1\), the same parameter increase shows an effect that increases the spread and asymmetry of the distribution. The sign change observed in \(S_x\) is particularly noteworthy. This transition shows that the skewness of the distribution, and therefore which tail of the population-density distribution becomes dominant, is determined by the fear parameter.

These results show that the fear parameter \(f\) not only creates a scale shift but also functions as a control mechanism that reconstructs the mean, spread, and asymmetry of the distribution in different directions depending on the sign of the contact channel.

\subsection{Fisher information analysis}\label{subsec:fisher-information}

In stochastic dynamical systems, the stationary probability distribution characterises the long-time behaviour of the system. Fisher information provides a natural sensitivity measure for quantifying the effect of small changes in control parameters on this distribution. Fisher information for the steady state probability density \(P_{\mathrm{st}}(x|\theta)\) is defined as follows \cite{fisher1925}:
\begin{equation}
I(\theta)
=
\int_{0}^{\infty}
P_{\mathrm{st}}(x|\theta)
\left(
\partial_{\theta}\ln P_{\mathrm{st}}(x|\theta)
\right)^2
\,dx.
\label{eq:fisher_information}
\end{equation}

This quantity is the second moment of the derivative of the log-probability density with respect to the parameter under the distribution and measures the local sensitivity of the distribution in parameter space. In statistical estimation theory, Fisher information plays a fundamental role through the Cram\'er--Rao lower bound \(\mathrm{Var}(\hat{\theta})\geq1/I(\theta)\) and defines a natural Riemannian metric structure on the probability family in the multiparameter case \cite{fisher1925,cover2005}.

From a physical standpoint, Fisher information quantitatively characterises the structural sensitivity of the steady state probability distribution to small changes in the control parameter. In regions where \(I(\theta)\) takes large values, small perturbations in the parameter lead to significant changes in the shape of the distribution. This indicates that the distribution is highly sensitive to the parameter. In contrast, in regions where \(I(\theta)\) is small, the distribution responds relatively weakly to parameter changes and points to a more structurally stable regime. In this context, Fisher information can be interpreted as a quantitative indicator of parametric sensitivity and the tendency towards structural reorganisation in stochastic systems.

\vspace{-0.3em}
\begin{figure}[h]
\centering
\includegraphics[width=0.95\textwidth]{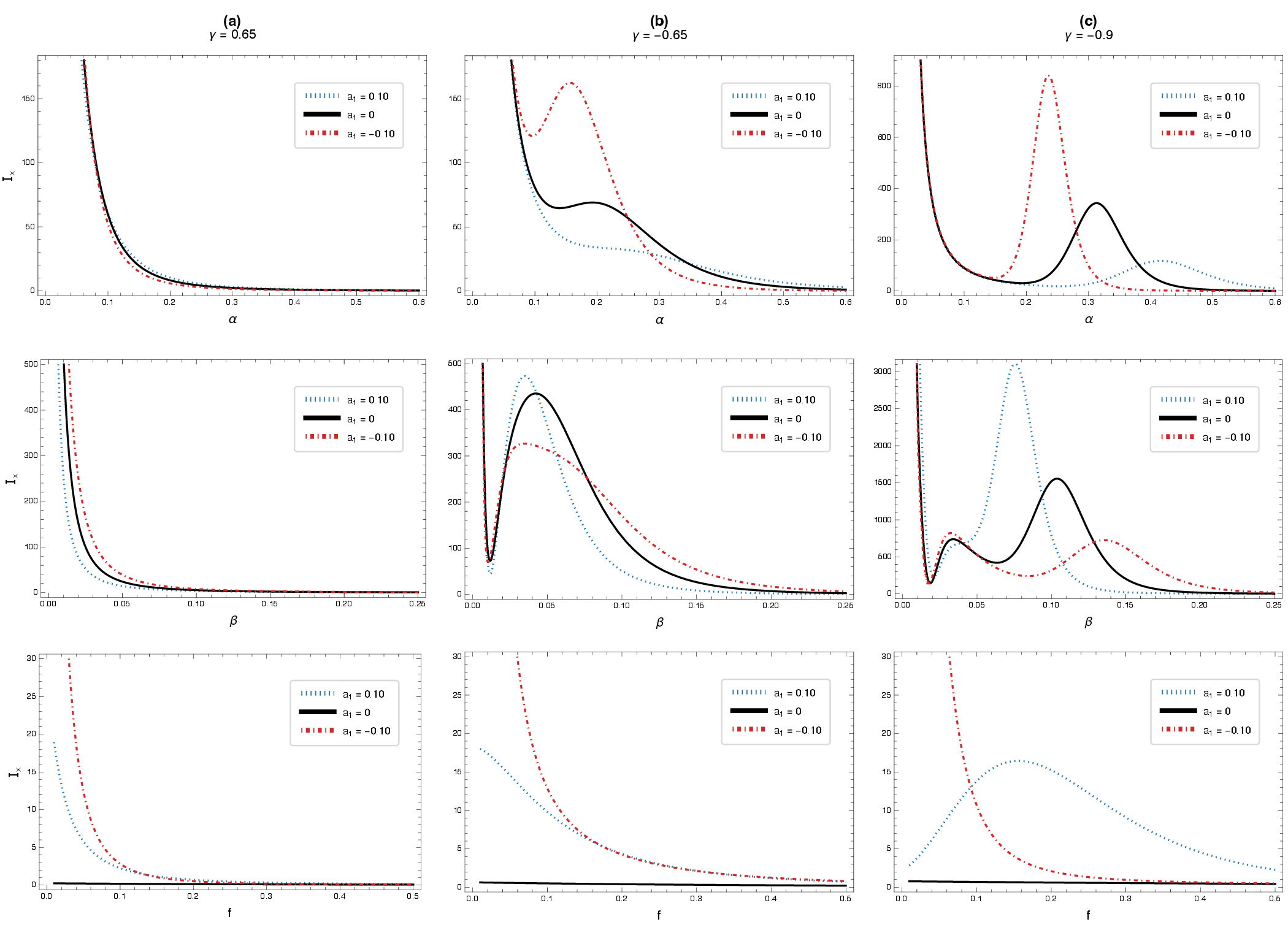}
\caption{Variation of Fisher information calculated from the steady state probability distribution \(P_{\mathrm{st}}(x|\theta)\) under systematic parameter scans. The rows represent scans performed along the parameters \(\alpha\), \(\beta\), and \(f\), respectively. The columns correspond to the correlation coefficient values (a) \(\gamma=0.65\), (b) \(\gamma=-0.65\), and (c) \(\gamma=-0.9\). In all panels, \(r=1\), \(m=2\), \(K=4.5\), \(v=1\), and \(a_2=f\). In the first row, systematic scanning was performed along \(\alpha\) with fixed values of \(\beta=0.085\) and \(f=0.5\); in the second row, along \(\beta\) with fixed values of \(\alpha=0.45\) and \(f=0.5\); and in the third row, along \(f\) with fixed values of \(\alpha=0.45\) and \(\beta=0.085\).}
\label{fig:fig12}
\end{figure}

Figure~\ref{fig:fig12} shows the variation of Fisher information calculated from the steady state probability distribution \(P_{\mathrm{st}}(x|\theta)\) along \(\alpha\), \(\beta\), and \(f\). In the case of positive correlation \((\gamma=0.65)\), Fisher information decreases monotonically with the increasing parameter value. This indicates that the distribution becomes progressively less sensitive to parameter changes. In contrast, under negative correlation \((\gamma=-0.65\) and \(\gamma=-0.9)\), peaks appear in certain parameter ranges. These peaks show that the steady state distribution exhibits high sensitivity to small perturbations at the relevant parameter values and that the distribution structure is rapidly reshaped. Furthermore, the position and magnitude of the peaks shift depending on the sign of \(a_1\). Thus, different interaction scenarios directly affect the parametric sensitivity.

However, in the parameter ranges where Fisher information approaches zero, the term \(\partial_{\mathrm{\theta}}\ln P_{\mathrm{st}}(x|\theta)\) becomes negligible. This indicates that the dependence of the distribution on the parameter practically vanishes and that the distribution is structurally stabilised against small parameter changes. Therefore, large values of Fisher information correspond to regimes where parametric sensitivity is concentrated, while very small values correspond to parametrically and structurally stable regimes.

Fisher information analysis shows that the sensitivity of the steady state distribution to parameter changes manifests differently depending on the scenarios and noise correlation. While Fisher information generally tends to decrease under positive correlation, peak structures appear in certain parameter ranges under negative correlation. These peak structures can be associated with regions where the distribution is more rapidly reorganised in response to parameter changes. Thus, while moment analysis reveals the geometric restructuring of the distribution, Fisher information provides a complementary perspective, showing in which regions these transformations become more pronounced in terms of parametric sensitivity.

\section{Mean first-passage time analysis}\label{sec:mfpt}

The steady state analysis performed in the previous sections characterises the long-time-limit behaviour of the system, but it does not provide information on how long the transitions between steady states take. In this section, the mean first-passage time (MFPT) is examined to characterise the temporal dynamics of the system. MFPT is defined as the average time it takes for a stochastic system to reach a predefined target point for the first time, starting from a certain initial state \cite{gardiner1997,redner2001}. In this study, the average time it takes for the system to reach the unstable equilibrium point, or threshold value, \(x_{s_1}\), which forms the left escape threshold of the high-density basin, starting from the high-density stable equilibrium point \(x_{s_2}\), is calculated. A small MFPT value indicates that the system can easily exit the HD state and leave the high-density basin. A large MFPT value indicates that the high-density basin has strong attraction and that the population can remain in this state for a long time. The equilibrium points \(x_{s_1}\) and \(x_{s_2}\) are obtained from the numerical solution of the condition \(dx/dt=0\) for deterministic dynamics.

The MFPT function \(T(x)\), which gives the average time required to reach the target point from the starting point \(x\), satisfies the following backward Fokker--Planck equation \cite{gardiner1997}:

\begin{equation}
A(x)\frac{dT(x)}{dx}
+
B(x)\frac{d^2T(x)}{dx^2}
=
-1.
\label{eq:backward_fp}
\end{equation}
\vspace{0.05em}

Here, \(A(x)\) represents the drift term and \(B(x)\) represents the diffusion term. Suitable boundary conditions for the escape problem are taken as follows:
\begin{equation}
T(x_{s_1})=0,
\qquad
\frac{dT(x_{s_2})}{dx}=0.
\label{eq:mfpt_boundary_conditions}
\end{equation}

The first condition states that the passage time is zero when the target point \(x_{s_1}\) is reached, corresponding to an absorbing boundary. The second condition states that the probability flux is zero at the starting point \(x_{s_2}\), corresponding to a reflecting boundary. Under the assumption of zero probability flux in the steady state, the steady state probability distribution can be written as
\vspace{0.4em}
\begin{equation}
P_{\mathrm{st}}(x)
=
\frac{N}{B(x)}
\exp\left[
\int_{0}^{x}\frac{A(x')}{B(x')}\,dx'
\right]
=
\frac{N}{\sqrt{B(x)}}
\exp\left[
-\frac{U(x)}{\beta}
\right].
\label{eq:sspd_effective_potential}
\end{equation}

From this, the effective potential is found as
\begin{equation}
U(x)
=
\frac{\beta}{2}\ln |B(x)|
-
\beta
\int_{0}^{x}
\frac{A(x')}{B(x')}\,dx'.
\label{eq:effective_potential_mfpt}
\end{equation}

Under the boundary conditions in Eq.~\eqref{eq:mfpt_boundary_conditions}, the integral representation of the MFPT is given by
\begin{equation}
T(x_{s_2}\to x_{s_1})
=
\int_{x_{s_2}}^{x_{s_1}}
\frac{dx}{B(x)P_{\mathrm{st}}(x)}
\int_{0}^{x}
P_{\mathrm{st}}(y)\,dy.
\label{eq:mfpt_integral}
\end{equation}

In the small-noise limit, this expression is approximately reduced to a Kramers-type exponential form as follows \cite{gardiner1997,kramers1940}:
\begin{equation}
T(x_{s_2}\to x_{s_1})
\approx
\frac{2\pi}
{\sqrt{\left|V''(x_{s_2}) V''(x_{s_1})\right|}}
\exp\left[
\frac{U(x_{s_1})-U(x_{s_2})}{\beta}
\right].
\label{eq:kramers_mfpt}
\end{equation}

Thus, when Eq.~\eqref{eq:deterministic_potential} and Eq.~\eqref{eq:effective_potential_mfpt} are substituted into Eq.~\eqref{eq:kramers_mfpt}, the following explicit expression is obtained:
\begin{equation}
\begin{aligned}
T
\approx
&
\left|
\left(
\frac{2a_1x_{s_2}}{\left(1+a_2x_{s_2}^{2}\right)^2}
+
\frac{Kmr-2(K+m)rx_{s_2}+3rx_{s_2}^{2}}
{K(1+fv)}
\right)
\right.\\
&\left.
\times
\left(
\frac{2a_1x_{s_1}}{\left(1+a_2x_{s_1}^{2}\right)^2}
+
\frac{Kmr-2(K+m)rx_{s_1}+3rx_{s_1}^{2}}
{K(1+fv)}
\right)
\right|^{-1/2}
\\
&\times
\left(
\frac{
2\pi\sqrt{\beta x_{s_1}^{2}+2x_{s_1}\gamma\sqrt{\alpha\beta}+\alpha}
}{
\sqrt{\beta x_{s_2}^{2}+2x_{s_2}\gamma\sqrt{\alpha\beta}+\alpha}
}
\right)
\exp\left[
\Lambda(x_{s_2})-\Lambda(x_{s_1})
\right].
\end{aligned}
\label{eq:explicit_mfpt}
\end{equation}

Here, \(T\), whose explicit dependence on the system parameters and noise parameters is given, represents the mean first-passage time required for the system to reach the threshold, or unstable, point \(x_{s_1}\) from the high-density stable equilibrium point \(x_{s_2}\).

\begin{figure}[H]

\centering

\includegraphics[width=0.98\textwidth]{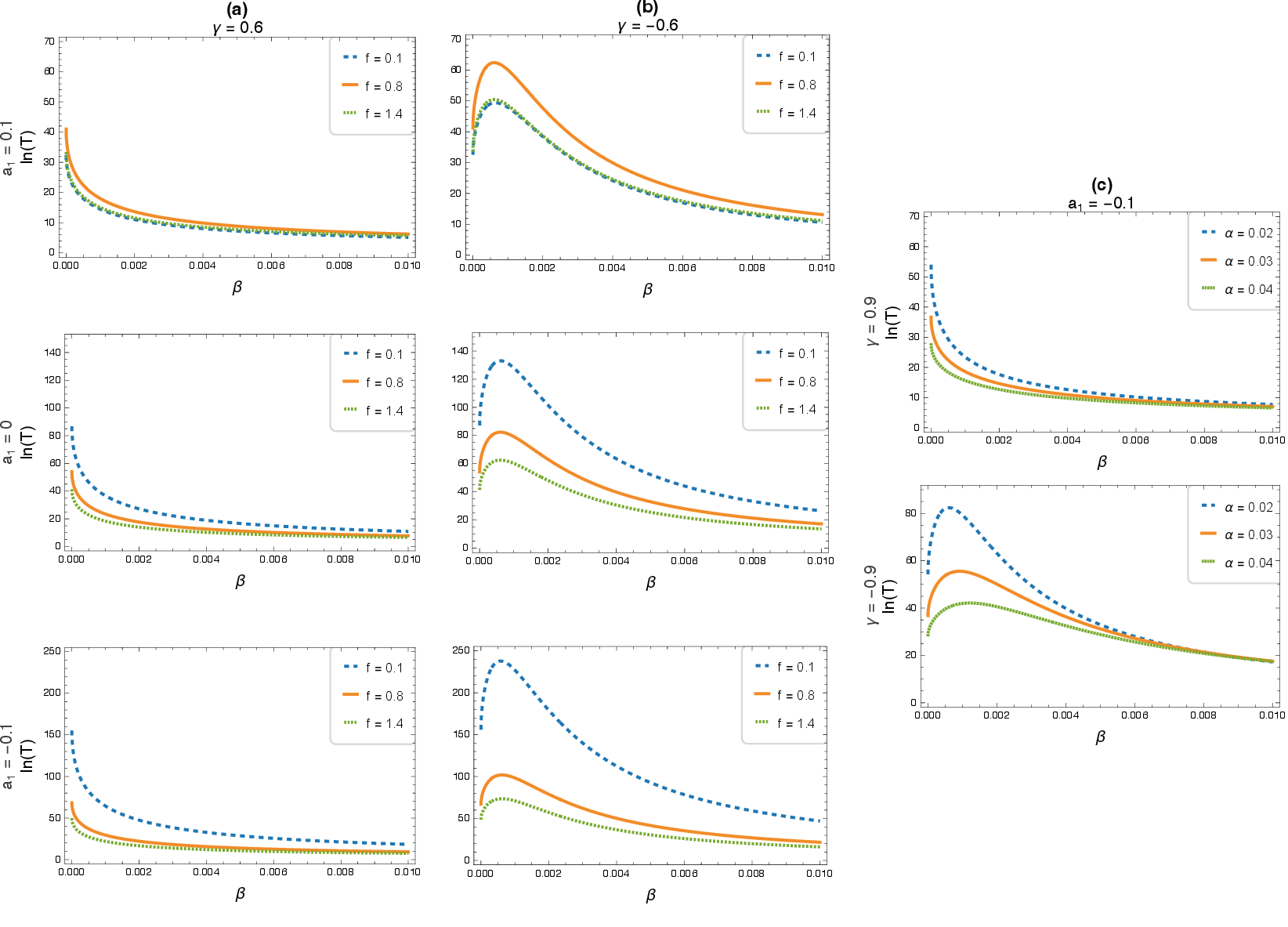}

\caption{Variation of \(\ln(T)\) as a function of the multiplicative noise intensity \(\beta\), with parameters \(r=1\), \(K=4.5\), \(m=2\), \(v=1\), \(\alpha=0.02\), and \(a_2=f\). (a) For \(\gamma=0.6\), from top to bottom, \(a_1=0.1\), \(a_1=0\), and \(a_1=-0.1\), respectively. (b) For \(\gamma=-0.6\), from top to bottom, \(a_1=0.1\), \(a_1=0\), and \(a_1=-0.1\), respectively. (c) For \(a_1=-0.1\), \(f=1.4\), from top to bottom, \(\gamma=0.9\) and \(\gamma=-0.9\), respectively.}

\label{fig:fig13}

\end{figure}

\vspace{-0.7em}

Figure~\ref{fig:fig13} shows the variation of MFPT on a logarithmic scale \((\ln T)\) depending on the multiplicative noise intensity \(\beta\). Figures~\ref{fig:fig13}a and \ref{fig:fig13}b represent different scenarios with \(a_1=0.1\), \(a_1=0\), and \(a_1=-0.1\) for positive \((\gamma=0.6)\) and negative \((\gamma=-0.6)\) correlation between noises, respectively. The curves correspond to different values of the fear parameter \(f\). In Figure~\ref{fig:fig13}c, the effect of the additive noise intensity \(\alpha\) for \(a_1=-0.1\) is compared under strong positive and negative correlations.

In the case of positive correlation in Figure~\ref{fig:fig13}a, MFPT decreases monotonically as \(\beta\) increases for all scenarios. This decrease is sharper in the low-\(\beta\) region and reaches saturation at high \(\beta\) values. Under positive correlation, additive and multiplicative noise are coupled with the same sign. This in-phase interaction increases the effectiveness of noise-induced fluctuations, facilitating barrier crossing and leading to a monotonic decrease of MFPT with \(\beta\).

The transition of the parameter \(a_1\) to negative values represents the change of the Holling type III term from a contact-induced loss character to a limited contact-induced demographic contribution character. This change strengthens the effective potential barrier by restructuring the deterministic drift term in favour of the high-density (HD) regime. Therefore, the mean first-passage time required for the system to leave the HD state increases significantly.

In contrast, in the \(a_1=0.1\) scenario, the Holling type III term represents contact-induced demographic loss. In this case, the fear parameter \(f\)'s growth-suppressive effect and the contact-induced loss mechanism reorganise the deterministic drift through two different channels. At low \(\beta\) values, this interaction tends to weaken the HD well, while as \(\beta\) increases, the interaction between multiplicative noise and drift becomes dominant at different scales. Therefore, the effective barrier height does not show a unidirectional change with respect to \(f\). As a result of the competing dynamics created by fear through two channels, MFPT curves belonging to different \(f\) values can intersect by switching places in certain \(\beta\) ranges. Since suppression increases in one channel while the saturation effect comes into play in the other channel, the net result varies depending on the magnitude of \(\beta\). For example, the intersection of the \(f=0.1\) and \(f=1.4\) curves in the case of \(a_1=0.1\) in Figure~\ref{fig:fig13}a is an indication of this. This behaviour is qualitatively different from the more regular and largely ordered MFPT structure observed in the \(a_1<0\) scenario.

As shown in Figure~\ref{fig:fig13}b, the behaviour of MFPT changes qualitatively under negative correlation. In all scenarios, a distinct maximum appears in the small-\(\beta\) region. This peak structure can be explained by the transient increase of the effective potential barrier in a certain \(\beta\) range as a result of the coupling of additive and multiplicative noise channels in opposite phase. At large \(\beta\) values, diffusion becomes dominant, the barrier is rapidly eroded, and MFPT decreases again. Consequently, the sign of the correlation creates not only a quantitative difference but also a qualitative transformation in the transition dynamics.

Figure~\ref{fig:fig13}c examines the effect of additive noise intensity \(\alpha\) on MFPT under strong correlation. While the monotonic decrease of MFPT with \(\beta\) is preserved under strong positive correlation, a distinct peak structure emerges under strong negative correlation. With increasing \(\alpha\), the peak height decreases and its position shifts towards larger \(\beta\) values. This behaviour shows that additive noise directly affects the modulation of the effective potential barrier together with correlation and redefines the scale of the transition dynamics.

Overall, MFPT analysis shows that the stability of the high-density population state is determined not only by the deterministic potential barrier, but also by the noise intensities, inter-noise correlation, and the interaction of the fear parameter across the two channels. MFPT exhibits parameter-sensitive and sometimes non-monotonic behaviour that cannot be explained solely by the increase in noise intensities. Thus, the fear effect, Holling interaction, and inter-noise correlation together create rich and qualitatively different transition scenarios.

\newpage

\section{Conclusion}\label{sec:conclusions}

In this study, a novel population model is proposed for a single-species prey population under the Allee effect, in which the fear effect is incorporated into the system via two distinct functional channels, and the stochastic extension of this model is analysed. The fundamental novelty of the model is that the fear parameter is defined not only as a stress factor that suppresses the intrinsic growth rate, but also as an active modulation mechanism that rescales the saturation structure of the Holling type III term representing predator--prey interaction. Unlike the single-channel fear models common in the literature, this approach allows fear to have a simultaneous and structural effect on population dynamics through two channels. The model also reveals two different ecological scenarios depending on the sign of the interaction term. The interaction term represents contact-mediated demographic loss when the coefficient \(a_1>0\), and a limited demographic contribution associated with defensive behaviours when \(a_1<0\). Thus, not only the negative demographic effects of fear, but also the behavioural advantages that may arise under certain conditions are included in the dynamical framework.

At the deterministic level, this dual-channel mechanism directly alters the velocity-field topology and equilibrium structure of the system. While in classical single-channel models equilibrium points are determined by the Allee threshold, in the proposed model the fear parameter becomes an independent control mechanism, rearranging the equilibrium topology and changing the position of the equilibrium points and the steady states to which the trajectories are directed. This deterministic restructuring forms the basis of the rich dynamics that emerge in the stochastic extension.

Within the scope of the stochastic analysis of the model, the steady state probability distribution \(P_{\mathrm{st}}(x)\) was analytically obtained using the Fokker--Planck formalism derived from the Langevin equation corresponding to the system. The accuracy of the analytical solution was tested with numerical simulations performed using the Milstein method, and a high level of agreement was observed between the two approaches. Systematically examining the analytical distribution over different parameters reveals how the distribution reorganises in parameter space. In particular, three-dimensional \(P_{\mathrm{st}}(x)\) surfaces allow the probability redistribution between low-density (LD) and high-density (HD) regions to be visualised over parameter changes on a global scale. In addition, P-bifurcation diagrams obtained from the extremum condition of the stationary distribution make it possible to quantitatively determine how the number and position of the peaks of the distribution change with the parameters. The combined use of these two approaches enabled the analysis of both the geometric and topological structure of the stochastic regime organisation of the system.

As a result of these analyses, two different transition mechanisms were distinguished in the system. In the first case, the number of peaks in the distribution remains unchanged, only the probability weights are redistributed between the LD and HD regions, and this situation is defined as a regime shift. In the second case, the number of peaks in the distribution changes, transforming the double-peaked structure into a single-peaked structure, and this qualitative transformation is called a phase-regime shift. In our study, both types of transitions were observed, and it was shown that the transitions can be triggered not only by noise parameters but also by the fear parameter. In the scans performed along the fear parameter, significant transformations in the distribution structure appear even though the noise intensities are kept constant, and this indicates the presence of parameter-induced transitions in the system.

The stochastic regime robustness differs significantly in the two scenarios defined depending on the sign of the interaction term. In Scenario I \((a_1>0)\), the Holling type III term represents contact-induced demographic loss, making the high-density regime structurally more fragile, and as multiplicative noise increases, the HD well disappears at lower noise levels. In contrast, in Scenario II \((a_1<0)\), the contact channel produces a limited demographic contribution, thus supporting the high-density regime and preserving the double-peaked structure over a wider parameter range. This result demonstrates that the sign of the interaction channel is one of the fundamental mechanisms determining the stochastic phase robustness of the system.

Analyses performed along the fear parameter reveal a significant structural feature of the system. The fact that fear acts through two channels, both as a stress factor that suppresses growth and as a mechanism that modulates Holling type III saturation, generates structural competition within the system. Therefore, the distribution in scans along the fear parameter does not exhibit a simple peak-shift behaviour. Instead, a more complex and non-monotonic evolution emerges, where low- and high-density peaks weaken together at specific intervals, and the probability mass spreads towards the intermediate-density region. This behaviour is qualitatively different from the simpler regime shifts observed in single-channel fear models and demonstrates that the proposed dual-channel mechanism transforms stochastic regime organisation into a richer structure.

Moment analysis, performed to examine the statistical structure of the steady state distribution, reveals that the fear parameter has scenario-sensitive effects on the mean, spread, and asymmetry of the distribution. Normalised moments show that fear is not only a parameter that changes the population scale, but is also associated with structural changes in the shape of the distribution. In particular, the sign change observed along the fear parameter in the normalised third moment \(S_x\) shows that the direction of skewness of the distribution can change. This result reveals that transitions between low- and high-density regions are related not only to changes in the average population size, but also to transformations in the asymmetric structure of the distribution.

Fisher information analysis shows that the sensitivity of the steady state distribution to parameter changes manifests differently depending on the scenarios and noise correlation. While Fisher information generally tends to decrease under positive correlation, peak structures appear in certain parameter ranges under negative correlation. These peak structures can be associated with regions where the distribution is more rapidly reorganised in response to parameter changes. Thus, while moment analysis reveals the geometric restructuring of the distribution, Fisher information provides a complementary perspective, showing in which regions these transformations become more pronounced in terms of parametric sensitivity.

Within the scope of MFPT analysis, a Kramers-type analytical expression was obtained for the mean first-passage time. The monotonic decrease in MFPT with multiplicative noise under positive correlation is consistent with the facilitation of barrier crossing by in-phase coupling, while the appearance of peaks in small noise intervals under negative correlation indicates that opposite-phase coupling temporarily strengthens the barrier. The intersection of MFPT curves for different fear values confirms that the two-channel interaction can modulate the passage times in a non-monotonic manner. These results reveal that the persistence of the high-density state is determined not only by the deterministic potential barrier but also by the noise intensities, inter-noise correlation, and the two-channel interaction of the fear parameter.

This study includes more general physical implications for regime organisation in stochastic systems with multi-channel control parameters, beyond ecological applications. In classical Kramers escape theory, the mean first-passage time is determined by the crossing of a fixed effective potential barrier by noise, and the ratio between barrier height and noise intensity exponentially controls this time. In contrast, in the proposed model, the fear parameter \(f\) is coupled to the drift term through two competing channels, dynamically and parameter-dependently restructuring the effective potential landscape. This requires that transitions be considered not only as rare events triggered by diffusion, but also as a process in which the barrier structure itself evolves through the control parameter. The control parameter \(f\) determines not only the mean first-passage times, but also the topological structure of the steady state probability distribution, the P-bifurcation thresholds, and the distribution of Fisher information in parameter space simultaneously. Therefore, the regime organisation observed here cannot be explained by purely noise-induced Kramers escape occurring under a constant potential, nor solely by deterministic bifurcations. This framework proposes a more general physical perspective on the nature of transitions in stochastic dynamical systems, going beyond classical single-channel approaches.

However, the study has some limitations. In the model, environmental uncertainties are represented by Gaussian white noise. While this approach allows for the construction of analytically solvable stochastic models, white noise is an idealised representation of environmental fluctuations. Therefore, the results obtained here offer at least a phenomenological framework. Examining models that consider noise processes with finite correlation times may allow for more realistic assessments in the future.

Studies reporting that predator risk arises through independent physiological mechanisms or that an increase in prey population size is observed with increasing risk at certain scales \cite{mao2005} point to the inadequacy of representing fear with a single suppression mechanism. Our model offers a suitable explanation for such contradictory empirical results at certain scales and quantitatively demonstrates that behavioural stress can also shape more subtle dynamical features. Of course, the fact that the model parameters have not yet been calibrated with direct empirical data is another limitation of the study. However, such a calibration process in ecological systems is inherently quite difficult, as frequently emphasised in the literature. Although the direct killing, or consumption, effects of predators on prey populations can be easily observed, the indirect costs of non-consumptive effects, such as fear on reproduction and survival, can only be measured indirectly through inference. Despite these empirical challenges, determining the relative effects of the fear parameter on growth suppression and interaction saturation using field data would be significant for evaluating the applicability of the model to ecological systems.

The model can be expanded in different directions in the future. Including alternative mathematical forms of the fear function and mechanisms such as time delays and memory effects in the model could contribute to a more realistic representation of the role of fear in demographic processes. In addition, considering prey refuges, spatial heterogeneity, and diffusion processes would allow for a more comprehensive examination of the role of the fear effect in spatial and demographic dynamics. Such extensions could contribute to a better understanding of how the dual-channel fear mechanism operates in different ecological contexts.

Overall, this study demonstrates that the fear effect in stochastic population dynamics can be considered not only as an individual-level stress response, but also as a structural modulation mechanism capable of reorganising the potential landscape, statistical structure, and regime organisation of the system. The interaction between noise intensity, inter-noise correlation, and the fear parameter reveals that both noise-induced and parameter-induced transitions can occur in the system, and that this interaction can generate a multi-layered regime organisation in population dynamics.

These findings demonstrate that behavioural factors can influence not only the average population size but also dynamical properties such as regime resilience, distribution asymmetry, and transition timescales in stochastic ecological models. In this context, the proposed model, especially when MFPT analysis and distribution-based sensitivity measures are considered together, offers an analytical basis for conservation biology and ecosystem management studies in terms of understanding the critical transition dynamics of populations under stochastic environmental conditions.

\clearpage

\begin{appendices}

\section{Explicit form of the SSPD}
\label{app:sspd-explicit}

The explicit steady state probability distribution used in Eq.~\eqref{eq:sspd_compact} is
\begin{equation}
P_{\mathrm{st}}(x)
=
\frac{N e^{\Lambda(x)}}
{\beta x^2+2\gamma\sqrt{\alpha\beta}\,x+\alpha},
\label{eq:appendix_pst}
\end{equation}
with the normalisation condition
\begin{equation}
\int_{0}^{\infty}P_{\mathrm{st}}(x)\,dx=1.
\label{eq:appendix_normalisation}
\end{equation}

The exponent function is given by
{\small
\begin{align}
\Lambda(x)
={}&
\frac{
2\mu_2
\left[
\ln(\alpha)
-
\ln\left(\alpha+x\left(\beta x+2\gamma\sqrt{\alpha\beta}\right)\right)
\right]
+
rxn\beta
\left(
K\beta+m\beta+2\gamma\sqrt{\alpha\beta}-\frac{x\beta}{2}
\right)
}
{Kn\beta^3(1+fv)}
\nonumber\\[0.6em]
&+
\frac{
a_1(a_2\alpha-\beta)\arctan\left(x\sqrt{a_2}\right)
-
a_1\gamma\sqrt{\alpha\beta a_2}\ln\left(a_2x^2+1\right)
}
{n\sqrt{a_2}}
\nonumber\\[0.6em]
&+
\frac{
\mu_1
\left[
\arctan\left(
\frac{x\beta+\gamma\sqrt{\alpha\beta}}
{\sqrt{\alpha\beta(1-\gamma^2)}}
\right)
-
\arctan\left(
\frac{\gamma\sqrt{\alpha\beta}}
{\sqrt{\alpha\beta(1-\gamma^2)}}
\right)
\right]
}
{
Kn(1+fv)\beta^3\sqrt{\alpha\beta(1-\gamma^2)}
}.
\label{eq:appendix_lambda}
\end{align}
}

The auxiliary quantities are
\begin{equation}
n
=
a_2^2\alpha^2+\beta^2+2a_2\alpha\beta(2\gamma^2-1),
\label{eq:appendix_n}
\end{equation}
{\small
\begin{align}
\mu_1
={}&
a_2^2r
\Big[
Km(\alpha\beta)^{5/2}\gamma
+
\alpha^3(K+m)\beta^2(2\gamma^2-1)
+
\alpha^2(\alpha\beta)^{3/2}\gamma(4\gamma^2-3)
\Big]
\nonumber\\
&+
\beta^2
\Big[
Km\beta^2\gamma r\sqrt{\alpha\beta}
+
\alpha r(K+m)\beta^2(2\gamma^2-1)
+
(\alpha\beta)^{3/2}r\gamma(4\gamma^2-3)
\nonumber\\
&\hspace{4.5em}
+
a_1\alpha K\beta^2(1-2\gamma^2+fv-2fv\gamma^2)
\Big]
\nonumber\\
&+
a_2
\Big[
-a_1\alpha^2K\beta^3(fv+1)
+
6\gamma r(\alpha\beta)^{5/2}
-
20\gamma^3r(\alpha\beta)^{5/2}
+
8m\beta^2\gamma^4r(\alpha\beta)^2
\nonumber\\
&\hspace{4.5em}
+
16\gamma^5r(\alpha\beta)^{5/2}
+
\beta^3r\alpha^2(K+m)(2-8\gamma^2)
\nonumber\\
&\hspace{4.5em}
+
K\beta\gamma r
\left(
8(\alpha\beta)^2\gamma^3
+
m\beta(\alpha\beta)^{3/2}(4\gamma^2-2)
\right)
\Big],
\label{eq:appendix_mu1}
\end{align}
}
and
{\small
\begin{align}
\mu_2
={}&
\beta^3
\Bigg[
\frac{1}{2}mr\gamma\sqrt{\alpha\beta}
+
\frac{1}{4}K
\left(
mr\beta-\beta^2(1+fv)-2\gamma(a_1-r+a_1fv)\sqrt{\alpha\beta}
\right)
\nonumber\\
&\hspace{7em}
+
\alpha r\left(\gamma^2-\frac{1}{4}\right)
\Bigg]
\nonumber\\
&+
a_2^2\alpha
\Bigg[
\frac{1}{4}\alpha K\beta^2(mr-\beta(1+fv))
+
\frac{1}{2}\gamma r(K+m)(\alpha\beta)^{3/2}
+
\alpha^2r\beta\left(\gamma^2-\frac{1}{4}\right)
\Bigg]
\nonumber\\
&+
a_2
\Bigg[
\alpha^2r\beta^2\left(\frac{1}{2}-3\gamma^2\right)
+
r\gamma
\left(
4\alpha^2\beta^2\gamma^3
+
K\alpha^{3/2}\beta^{5/2}(2\gamma^2-1)
+
m\alpha^{3/2}\beta^{5/2}(2\gamma^2-1)
\right)
\nonumber\\
&\hspace{7em}
+
\alpha K\beta^3
\left(
mr\left(\gamma^2-\frac{1}{2}\right)
+
\frac{1}{2}\beta
-
\gamma^2\beta
+
fv\beta\left(\frac{1}{2}-\gamma^2\right)
\right)
\Bigg].
\label{eq:appendix_mu2}
\end{align}
}

\end{appendices}

\backmatter
\clearpage
\section*{Competing Interests}

The authors declare that they have no conflict of interest.

\section*{Authorship Contributions}

Ö.G. conceived the theoretical framework, carried out the analytical derivations, designed and implemented the numerical simulations, and wrote the original manuscript. M.M. carried out the numerical analyses, conducted parameter scans, produced the figures, and contributed to manuscript preparation. Both authors discussed the results and approved the final version of the manuscript.

\bibliography{sn-bibliography}

\end{document}